\documentclass{aa}  
\usepackage{graphicx}
\usepackage[varg]{txfonts}
\usepackage{natbib}
\bibpunct{(}{)}{;}{a}{}{,}
\begin{document} 

   \title{V838 Mon: A slow waking up of Sleeping Beauty?\thanks{Tables 2 and 4 are only available in electronic form at the CDS via anonymous ftp to cdsarc.cds.unistra.fr (130.79.128.5) or via https://cdsarc.cds.unistra.fr/cgi-bin/qcat?J/A+A/}}

   \author{T. Liimets 
          \inst{1,2}
          \and
          I. Kolka
          \inst{2}         
                \and
          M. Kraus 
          \inst{1} \and
         T. Eenm\"ae
          \inst{2} \and
          T. Tuvikene
          \inst{2} \and
          T. Augusteijn
          \inst{3, 4} \and
          L. Antunes Amaral
          \inst{5} \and
          A. A. Djupvik 
          \inst{3, 4} \and    
         J. H. Telting
          \inst{3, 4} \and  
          B. Deshev
          \inst{1} \and  
          E. Kankare  
          \inst{6} \and 
          J. Kankare 
           \and 
          J. E. Lindberg \and  
          T. M. Amby
          \inst{7} \and 
          T. Pursimo
          \inst{3, 4} \and 
          A. Somero
          \inst{6} \and 
          A. Thygesen 
          \and 
          P. A.~Str\o m
          \inst{8, 9}
           }

\institute{Astronomical Institute, Czech Academy of Sciences, Fri\v{c}ova 298, 25165 Ond\v{r}ejov, Czech Republic\\
              \email{tiina.liimets@asu.cas.cz}
         \and
               Tartu Observatory, University of Tartu, Observatooriumi 1, 61602 T\~oravere, Estonia    
         \and
         Nordic Optical Telescope, Rambla Jos\'{e} Ana Fern\'{a}ndez P\'{e}rez 7, ES-38711 Bre\~{n}a Baja, Spain 
         \and
         Department of Physics and Astronomy, Aarhus University, Munkegade 120, DK-8000 Aarhus C, Denmark 
         \and
         Instituto de F\'{\i}sica y Astronom\'{\i}a, Universidad de Valpara\'{\i}so, Gran Breta\~na 1111, Playa Ancha, Valpara\'{\i}so 2360102, Chile
          \and 
          Tuorla Observatory, Department of Physics and Astronomy, University of Turku, FI-20014 Turku, Finland    
          \and 
          Viborg Katedralskole, Gl.Skivevej 2, 8800 Viborg, Denmark
          \and
          Department of Physics, University of Warwick, Coventry CV4 7AL, UK 
          \and 
          Institut d'astrophysique de Paris, CNRS, UMR 7095 \& Sorbonne Universit\'es, UPMC Paris 6, $98^{\mathrm{bis}}$ Boulevard Arago, 75014 Paris, France
             }

   \date{Received ; accepted }

 
  \abstract
   {
   V838 Monocerotis is a peculiar binary that underwent an immense stellar explosion in 2002, leaving 
   behind an expanding cool supergiant and a hot B3V companion. Five years after 
   the outburst, the B3V companion disappeared from view, and has not returned to its original state.
   }  
   {We investigate the changes in the light curve and spectral features 
   to explain the behaviour of  V838 Mon during the current long-lasting minimum.}
   {A monitoring campaign has been performed over the past 13 years with the Nordic 
   Optical Telescope  to obtain optical photometric and spectroscopic data. The datasets are used to analyse the temporal evolution of the spectral features and the 
   spectral energy distribution, and to characterise the object.
   }
   {Our photometric data show a steady brightening in all bands over the past 13 years, 
   which is particularly prominent in the blue. This rise is also reflected in the spectra, 
   showing a gradual relative increase in the continuum flux at shorter wavelengths.
   In addition, a slow brightening of the H$\alpha$ emission line starting in 2015 was detected. 
     These changes might imply that the B3V companion is slowly reappearing.
     During the same time interval, our analysis reveals a considerable change in the observed 
   colours of the object along with a steady decrease in the strength and width of molecular 
   absorption bands in our low-resolution spectra.  
     These changes suggest a rising temperature of the cool supergiant along with a weakening of 
     its wind, most likely combined with a slow recovery of the secondary due to the evaporation 
     of the dust and accretion of the material from the shell in which the hot companion is embedded.     
  From our medium-resolution spectra, we find that the heliocentric radial velocity of the 
   atomic absorption line of Ti\,{\sc i} 6556.06\,\AA \ has been stable 
   for more than a decade. We propose that Ti\,{\sc i} lines are tracing the velocity 
   of the red supergiant in V838 Mon, and do not represent the infalling matter as previously stated.
}
   {}


   \maketitle
%

\section{Introduction}

\object{V838 Monocerotis} (V838 Mon) is a peculiar star discovered in January 2002 
\citep{2002IAUC.7785....1B} when it underwent a 
tremendous stellar explosion, creating a dense envelope of expanding molecular matter 
\citep{2011A&A...529A..48K, 2016A&A...596A..96E}. 
The eruption lasted three months, consisting of two or three maxima depending on the 
wavelength, and has been studied by many authors 
(e.g. \citealt{2002A&A...389L..51M,2003MNRAS.341..785C}, and references therein). 
The nature of the explosion is still uncertain. Currently, the most accepted model favours 
a stellar merger, placing V838 Mon in a rare class of objects, luminous red novae. 
The aftermath of the eruption was a cool L-type supergiant (SG) with a photospheric temperature 
between 2000 and 2300 K (\citealt{2003MNRAS.343.1054E,2006A&A...460..245P}).
The hot B3V secondary was 
discovered spectroscopically in October 2002 \citep{2002IAUC.7982....1D,2002IAUC.7992....2W}. 
Hence, the progenitor of the outburst could have been a triple (or  higher multiple) system.  

Shortly after the maximum brightness, a spectacular light echo occurred 
\citep{2002IAUC.7859....1H} that  was observable for several years. 
The light echo of V838 Mon is the most studied light echo
in the history of astronomy using ground-based observations (e.g. \citealt{2005MNRAS.358.1352C}, 
\citealt{2007ASPC..363..174L}, and references therein) 
and the Hubble Space Telescope \citep{2003Natur.422..405B,2007ASPC..363..130B}.

For a few years following the 2002 outburst V838 Mon displayed a stable light curve until November-December 2006, when it experienced an eclipse-like event 
\citep{2006ATel..964....1G,2006ATel..966....1B} that lasted about 70 days. 
At first, it was considered to be an eclipse of the hot B3V companion. 
However, the emission line spectrum, associated with the secondary,  
remained strong during this period (\citealt{2009A&A...503..899T}). 
Later on, it was shown by \cite{2009A&A...503..899T} that the brief dimming 
was caused by the interaction of the expelled matter from the 2002 outburst.  
The system recovered from the fading to its original 
brightness only to experience another deeper \citep{2008ATel.1821....1G} 
and still ongoing minimum.
The second decline took more than two years to reach the minimum brightness 
(\citealt{2020AstBu..75..325G}, hereafter GOR20, and our Fig.~\ref{F-lcour}), 
and during this period  the B3V star 
also disappeared from the spectrum. 
Large amounts of dust, formed in 2007-2008 from the matter expelled during the
2002 outburst, have reached the hot component (\citealt{2011A&A...532A.138T}). 
The interferometric observations show that V838 Mon is surrounded by an extended  disc-like dusty 
environment 
\citep{2014A&A...569L...3C, 
2020A&A...638A..17O, 2021A&A...655A..32K, 2021A&A...655A.100M}. Accretion from
this dust has led to the formation of a shell around the B3V star \citep{2021A&A...655A.100M}, 
which is most likely the cause of the continuing
occultation. The subsequent evolution of this dust shell is governed by
evaporation due to ionising radiation from the hot secondary, and by ongoing
accretion from the replenishing material supplied from the ejecta and the
wind of the primary.

In this paper we analyse data collected during the ongoing minimum, and shed 
light on the cause of the photometric and spectroscopic 
variability during these years. In Sect.~\ref{S-data} we present our data and the 
reduction procedures. Our results are presented in Sect.~\ref{S-results}, which are discussed 
in Sect.~\ref{S-disc}, and our conclusions are summarised in Sect.~\ref{S-conc}.


\section{Observations and data reduction}\label{S-data}

\subsection{Optical photometry}\label{S-phot}

The bases of this paper are the data collected 
since 2009 with the 2.56 m Nordic Optical 
Telescope\footnote{https://www.not.iac.es} 
(NOT) equipped with the cryogenically cooled 
Alhambra Faint Object Spectrograph and Camera\footnote{http://www.not.iac.es/instruments/alfosc} (ALFOSC). 
Bessel $UBVR$ filters 
and interference filter $i$ (NOT ID no. 12) with a central wavelength of 797 nm and FWHM 157 nm 
were exploited. The  Bessel filters are equivalent to the Johnson-Cousins $UBVR_C$ system. 
The measurements in ALFOSC $i$ filter were calibrated using the fixed 
colour terms provided at the ALFOSC website\footnote{http://www.not.iac.es/instruments/alfosc/zpmon/}, 
which were obtained using the Landolt 
standard stars in the standard $UBVR_CI_C$ system, and therefore take into account the 
transformation to $I_C$ system.
Observations were conducted through normal observing programmes\footnote{Proposals P39-410; P42-044; 
P44-042; P46-015; P48-030; P50-038; P52-032; P54-024; P56-025; P58-025}, and on 
2019 November 29 via the FINCA observing school.
Several measurements were obtained in each season (in earlier years once a month, in later years once every six months), 
with $B$ and $R_C$ bands being more frequent, 
approximately every two weeks. In total V838 Mon was observed on 100 nights 
with a  typical seeing of $1''.0$.
Our complementary photometric data, obtained with the telescopes at the 
Tartu Observatory (TO) and South African Astronomical Observatory (SAAO) between the years 2002 and 2019, 
are described in Appendix~\ref{A-phot}.

\begin{figure*}[!h]
\centering
\includegraphics[width=17cm]{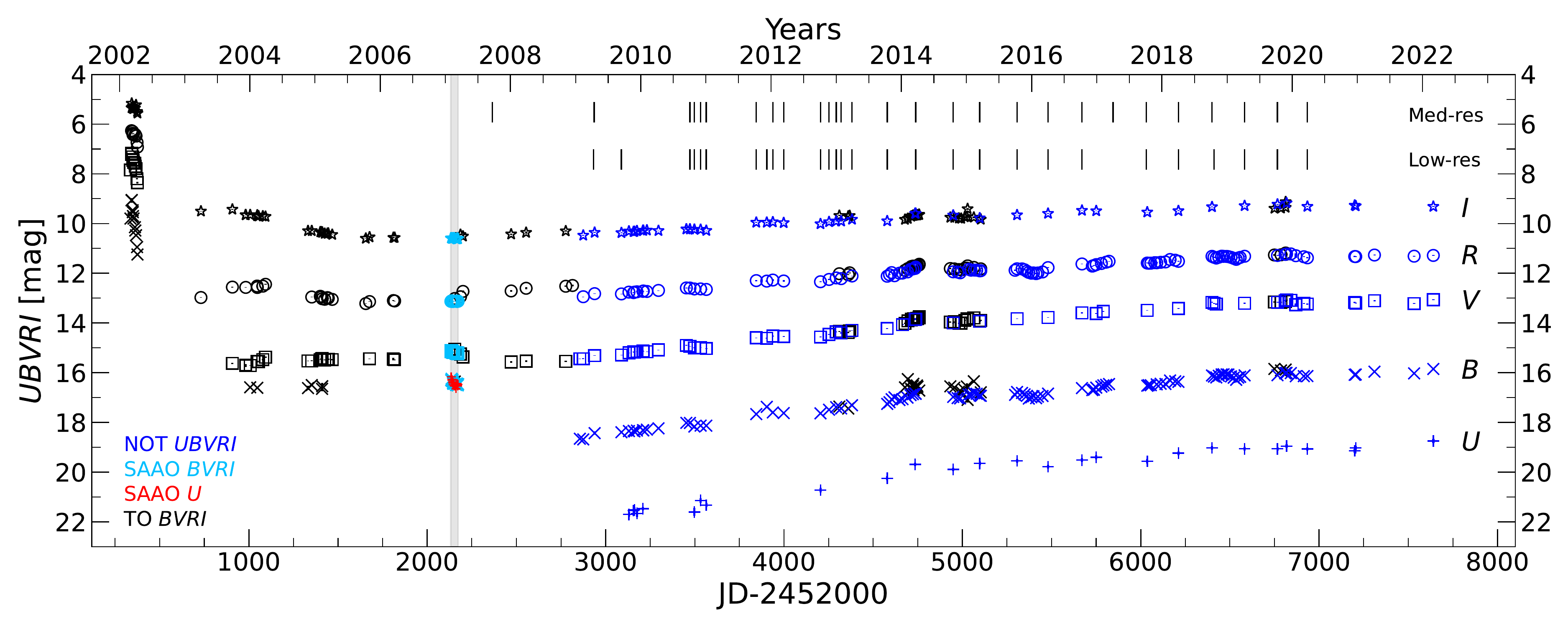} 
\caption{Photometric data of V838 Mon. Colour-coding refers to the data from the various observatories, 
and different symbols indicate the various photometric bands. The grey area shows the period of the SAAO 
observations in the beginning of 2007. The $U$-band measurements from SAAO are marked in red, 
to distinguish them from the $B$-band values of similar magnitude.
The vertical lines mark the dates with our low- and medium-resolution spectra.}
\label{F-lcour}
\end{figure*}

All the data were bias and flat-field corrected using either the Interactive Data Language (IDL) 
for data taken up to 2007 or (IRAF)\footnote{IRAF is distributed by the National 
Optical Astronomy Observatory, which is operated by the Association 
of Universities for Research in Astronomy (AURA) under cooperative 
agreement with the National Science Foundation.} for data collected afterwards. 
For the telescopes at TO dark current correction was also performed. 
For all the different datasets the magnitude measurements were done using aperture photometry,  
which provides reliable measurements during all states of 
V838 Mon (e.g. \citealt{2003MNRAS.341..785C,2007ASPC..363....3H}, GOR20).
We note  that our observations are not contaminated by the light echo.
The transformation from an instrumental system to a standard Johnson-Cousins photometric 
system was performed using the colour transformation 
coefficients obtained by observing Landold standard fields for each 
telescope\footnote{For  NOT the coefficients can be found 
at http://www.not.iac.es/instruments/alfosc/zpmon/} and comparison stars in the field of view (FOV), 
which were checked beforehand  for  possible variability,  a suspicion 
triggered by the photometric discrepancies mentioned in \cite{2007AJ....133..387A}.
For the variability check we cross-matched the standard magnitudes 
of  V838 Mon's comparison stars presented in 
the literature \citep{2002A&A...389L..51M,2004IBVS.5511....1G,2005A&A...434.1107M}
and we calculated our own standard magnitudes of the stars presented in \cite{2002A&A...389L..51M}.
We conclude that all the stars that we
could test of the latter are suitable comparison stars. The analyses could
not be performed 
for star $l$, which was out of our FOV, and
stars $n$ and $p$ were too faint in $U$ to give
reliable measurements from our data. In addition, due to the faintness
of the V838 Mon starting from 2008, we added to
our list three stars (designated as $r, s,$ and $t$), and checked their stable photometric nature. 
Table~\ref{T-cs} lists the standard magnitudes of all comparison
stars measured in this work. 
These objects, together
with the brighter ones from \cite{2002A&A...389L..51M}, were used to calibrate 
instrumental magnitudes of all data in this work. 
Mostly, more than four comparison stars were used in 
obtaining the standard magnitude of V838 Mon. 
The log of all the photometric data of V838 Mon 
is provided in the online table (see Table~\ref{T-phot} for 
representative lines), and the measurements are presented in Fig.~\ref{F-lcour}. 
The errors in Table~\ref{T-phot} for the TO and SAAO data are a combination of 
the Poissonian error of instrumental magnitudes and the uncertainties of the colour transformation. 
For our main dataset from  NOT the errors in the $BVRi$ are the standard deviation  of V838 Mon standard magnitudes
when using several comparison stars. 
The errors of instrumental magnitudes are 
negligible in these filters. 
In $U$ band the Poissonian noise of the instrumental magnitudes was also 
taken into account;  during the start of our NOT observations, when  
V838 Mon was as faint as 21 magnitudes, the noise was considerable. 

From here on, we refer to the filters in this work as $UBVRI$.

\begin{table*}
\begin{small}
\caption{Standard magnitudes of the comparison stars calculated in this work. 
RAJ2000 and DEJ2000 are from the  Vizier catalogue presented in \cite{2005A&A...434.1107M}. 
ID$_{M02}$ refers to the star nomenclature given by \cite{2002A&A...389L..51M},
while the ID is given in this work. }  
\label{T-cs}
\centering 
\begin{tabular}{lllllllll}
\hline\hline 
ID$_{M02}$ & ID &  RAJ2000 & DEJ2000 & $U \pm U_{\sigma} $ & $B \pm B_{\sigma} $ & $V \pm V_{\sigma} $ & $R_C \pm R_{C\sigma} $ & $I_C \pm I_{C\sigma} $ \\
\hline
\textit{f} && 106.024244 & -3.844027 & 14.652 $\pm$ 0.045 & 14.670 $\pm$ 0.012 & 14.121 $\pm$ 0.006 & 13.776 $\pm$ 0.006 & 13.411 $\pm$ 0.009 \\ 
\textit{g} && 106.010135 & -3.881230 & 15.173 $\pm$ 0.037 & 14.975 $\pm$ 0.011 & 14.613 $\pm$ 0.006 & 14.413 $\pm$ 0.005 & 14.139 $\pm$ 0.008 \\ 
\textit{h} && 106.004135 & -3.874533 & 15.486 $\pm$ 0.048 & 15.421 $\pm$ 0.013 & 14.800 $\pm$ 0.007 & 14.452 $\pm$ 0.006 & 14.111 $\pm$ 0.009 \\ 
\textit{j} && 106.031128 & -3.851698 & 15.426 $\pm$ 0.046 & 15.618 $\pm$ 0.012 & 15.050 $\pm$ 0.006 & 14.676 $\pm$ 0.006 & 14.239 $\pm$ 0.009 \\ 
\textit{k} && 106.021087 & -3.810408 & 16.760 $\pm$ 0.022 & 16.354 $\pm$ 0.015 & 15.524 $\pm$ 0.008 & 15.033 $\pm$ 0.007 & 14.585 $\pm$ 0.011 \\ 
\textit{m} && 106.004593 & -3.842428 & 17.481 $\pm$ 0.056 & 17.325 $\pm$ 0.015 & 16.551 $\pm$ 0.008 & 16.074 $\pm$ 0.008 & 15.630 $\pm$ 0.013 \\ 
\textit{n} && 106.003805 & -3.862652 & --- & 18.650 $\pm$ 0.018 & 17.803 $\pm$ 0.012 & 17.309 $\pm$ 0.013 & 16.741 $\pm$ 0.027 \\ 
\textit{p} && 106.025250 & -3.820849 & --- & 19.374 $\pm$ 0.024 & 18.478 $\pm$ 0.015 & 17.909 $\pm$ 0.018 & 17.341 $\pm$ 0.039 \\ 
        &\textit{r}& 106.058063 & -3.888445 & 15.735 $\pm$ 0.048 & 15.641 $\pm$ 0.013 & 15.020 $\pm$ 0.007 & 14.630 $\pm$ 0.006 & 14.240 $\pm$ 0.009 \\ 
         &\textit{s}& 105.981531 & -3.834276 & 18.060 $\pm$ 0.057 & 17.748 $\pm$ 0.014 & 17.058 $\pm$ 0.008 & 16.631 $\pm$ 0.009 & 16.105 $\pm$ 0.016 \\ 
         &\textit{t}& 106.058678 & -3.866647 & 17.964 $\pm$ 0.056 & 17.786 $\pm$ 0.014 & 17.111 $\pm$ 0.008 & 16.654 $\pm$ 0.009 & 16.207 $\pm$ 0.017 \\ 
\hline 
\end{tabular}
\end{small}
\end{table*}

\begin{table*}[!t]

\caption{Journal of photometric observations and their uncertainties. 
The entire table is available in electronic form at the CDS. Here only the first five lines are presented for guidance.
Instrument names are explained in the text.}              
\label{T-phot}      
\centering                                      
\begin{tabular}{c|ll|ll|ll|ll|ll|r}          
\hline\hline                        
JD &   $U$ & $\sigma_{U}$   &   B  & $\sigma_{B}$   &    V& $\sigma_{V}$ &     R& $\sigma_{R}$   &   I & $\sigma_{I}$   & Instrument \\
\hline 
2452336.30  &   $-$ &  $-$ & 9.798 &  0.025 &  7.843 &  0.018 & $-$ &  $-$  & $-$ &  $-$  & HPC \\
2452342.32  &   $-$ &  $-$ & 9.066 &  0.087 &  7.211 &  0.021 & 6.263  & 0.018    & 5.347  & 0.032    & HPC \\
2452343.26  &   $-$ &  $-$ & 9.051 &  0.026 &  7.173 &  0.018 & $-$ & $-$   & 5.166  & 0.024    & HPC \\
2452348.39  &   $-$ &  $-$ & 9.465 &  0.023 &  7.316 &  0.029 & 6.293  & 0.015    & 5.291  & 0.038    & HPC \\
2452349.32  &   $-$ &  $-$ & 9.541 &  0.007 &  7.423 &  0.016 & 6.367  & 0.016    & 5.347  & 0.023    & HPC \\
\hline                           
\hline        
\end{tabular}
\end{table*}

\subsection{Infared photometry}\label{S-irdata}

V838 Mon was observed with NOTCam\footnote{The Nordic Optical Telescope's near-IR Camera and spectrograph 
http://www.not.iac.es/instruments/notcam/}
on 2010 May 11 and 2020 March 2.
The high-resolution camera (0.079"/pix) was used to spread the flux over many
pixels, and imaging through the broad-band filters $J$, $H$, and $K_{s}$ was obtained. 
Due to the brightness of V838 Mon in 2010, the shortest recommended exposure
times were used along with some defocusing, most notably for $H$ and $K_{s}$. 
The FOV contained three 2MASS stars that were used for calibration. 
In 2020 the target was even brighter and 
we added a small cold stop in the beam  to reduce the transmission to roughly 20\%. 
In this case the 
field stars became too faint for calibrations, and we observed a
standard star field at a similar airmass just before the target, 
using the same stop. Reduction of the dithered images was
done with the notcam.cl IRAF package, following a standard 
procedure with flat-fielding, sky-subtraction, shifting, and
adding of frames by median combination to exclude bad pixels. 
Photometry was done using large apertures, and the
error in the magnitudes is dominated by the uncertainty in the calibration stars. 
The analysis performed in this paper requires the $K$-band magnitudes in the 
\citet{1988PASP..100.1134B} photometric system. Therefore, we 
transformed our 2MASS calibrated $K_s$ magnitudes into $K_{BB}$ using 
the relations provided by the IRSA webpage\footnote{https://irsa.ipac.caltech.edu/data/2MASS/docs/releases/allsky/}.
Results are presented in Table~\ref{T-ir}. 
In all the following analyses the $K_{BB}$ magnitudes are used.

\begin{table}
\caption{Infrared photometric magnitudes. Dates are in the format of YYYY-MM-DD.}              
\label{T-ir}      
\centering                                      
\begin{tiny}
\begin{tabular}{l c c c c}          
\hline\hline                        
Date & $J$ & $H$ & $K_s$ & $K_{BB}$ \\    
&[mag]&[mag]&[mag]& [mag]\\
\hline                           
2010-05-11 &7.00 $\pm$ 0.06 & 5.86 $\pm$ 0.04 & 5.10  $\pm$ 0.11 & 5.14 $\pm$ 0.11 \\
2020-03-02\tablefootmark{a} & 6.59 $\pm$ 0.05 & 5.55 $\pm$ 0.05 & 4.76 $\pm$ 0.05 & 4.80 $\pm$ 0.05\\
\hline                                             
\end{tabular}
\tablefoot{
\tablefoottext{a}{Published in \cite{2021AJ....162..183W}} 
}
\end{tiny}
\end{table}

\subsection{Spectroscopy with the Nordic Optical Telescope}


Low-resolution long-slit spectra were regularly collected, with NOT and ALFOSC, 
on a total of 28 nights starting from 2009. 
The observations were typically carried out  twice a year, occasionally more frequently.
Full optical coverage was achieved with  grism 4 and a slit width of mostly  $1\farcs 0$  
providing spectral dispersion of $\sim$3 \AA\ pix$^{-1}$ (R $\sim$ 600 at 6600 \AA). 
Typical exposure times 
were 400 seconds which provides a S/N between $\sim$10  
and $\sim$100 over the usable spectral range \hbox{(4500 -- 9000 \AA )}.
In this way, the saturation of the red part of the cool star 
spectrum was avoided.

Additional medium-resolution spectra were acquired on 28 nights. 
These were taken around the \hbox{H$\alpha$ 6562.8 \AA\ } line with   grism 17 
and a slit width of mostly $0\farcs 5$. 
The spectra cover the range between 6330 and 6870 \AA\ and have a dispersion of 
\hbox{$\sim$ 0.26 \AA\ pix$^{-1}$} (R $\sim$ 9000 at 6600 \AA).
The exposure times  usually varied between 500 and 1200 seconds guaranteeing a S/N ratio $\geq$ 100. 
In addition, a spectrum from September 2007 with the similar set-up  is used in this work for comparison purposes.
All spectra obtained with  NOT were observed with the slit at the parallactic angle. 

The standard data reduction steps were performed with the
MIDAS software package in the context of long-slit spectra. 
The flux calibration is based on the matching of the synthetic photometry 
of the instrumental flux 
in the NOT \hbox{ALFOSC} $UBVRi$ bands to the quasi-simultaneous 
high-precision photometry (cf. Sects.~\ref{S-phot}, \ref{S-lc}, and \ref{S-sed}). 
For synthetic photometry and brightness zero-point fluxes we   used the database
provided by the Spanish Virtual Observatory\footnote{svo2.cab.inta-csic.es/theory/newov2/syph.php}. 
For clearer demonstration of the evolution of 
spectral shape (spectrophotometric gradient) of V838 Mon 
(cf. Sects.~\ref{S-lr} and \ref{S-hr}) we performed relative flux normalisation
using points selected in the stellar continuum 
at 7560 \AA\ and 6577  \AA , which 
we expect are close to the real continuum, for the grism 4 and 17 settings, respectively.
Telluric correction was omitted due to the strong contamination with the 
stellar molecular absorption bands.
The dates of the  spectral observations are indicated in Fig.~\ref{F-lcour}.
A complete log as well as the data can be acquired from the NOT FITS Header 
Archive\footnote{http://www.not.iac.es/observing/forms/fitsarchive/}. 
The representative example of reduced spectra is available in the CDS.


\section{Results}\label{S-results}

We note that most of the analyses in this article are based on  the data from  NOT 
in order to use a single telescope+instrument+filter set. This is important due 
to the extreme colours of V838 Mon during the past decade, which result in possible calibration difficulties 
(see Appendix~\ref{A-align}). 

\subsection{Light curve of V838 Mon}\label{S-lc}

The measurements of the full light curve in the different bands are presented in Table~\ref{T-phot} 
and are shown in Fig.~\ref{F-lcour}.
We note the lack of $U$-band measurements prior to 2007.
Our first $U$-band magnitudes are from early 2007 when the $U$ brightness  
was approximately equal to the $B$ magnitude. 
The results presented in this section also 
contain $V$ and $I$ measurements  previously presented by \cite{2007ASPC..363..174L}. 
In addition, we   added ALFOSC data presented by \cite{2009ATel.2211....1K}. 
We note that these had a calibration error resulting in an offset of less than $\sim$0.1 mag 
that we have corrected in the present work.

In Fig.~\ref{F-lcour} the light curve starts from early 2002 with the outburst (see also Fig.~\ref{F-lc2002B}). 
Then  a period of quiescence follows with rather stable magnitudes,
until late 2006 when a small eclipse-like event occurs (see Fig.~1 in  
\citealt{2007A&A...474..585M} and references 
 therein). This event was not covered by our observations and is therefore not 
visible in Fig.~\ref{F-lcour}.
The unexpected brightness drop of the system lasted 70 days 
(\citealt{2006ATel..964....1G}; \citealt{2006ATel..966....1B}) and was deeper in 
the bluer bands ($UBV$), while it remained almost unnoticed in $R$ and $I$. 

\begin{figure}[!t]
\centering
\includegraphics[width=9cm]{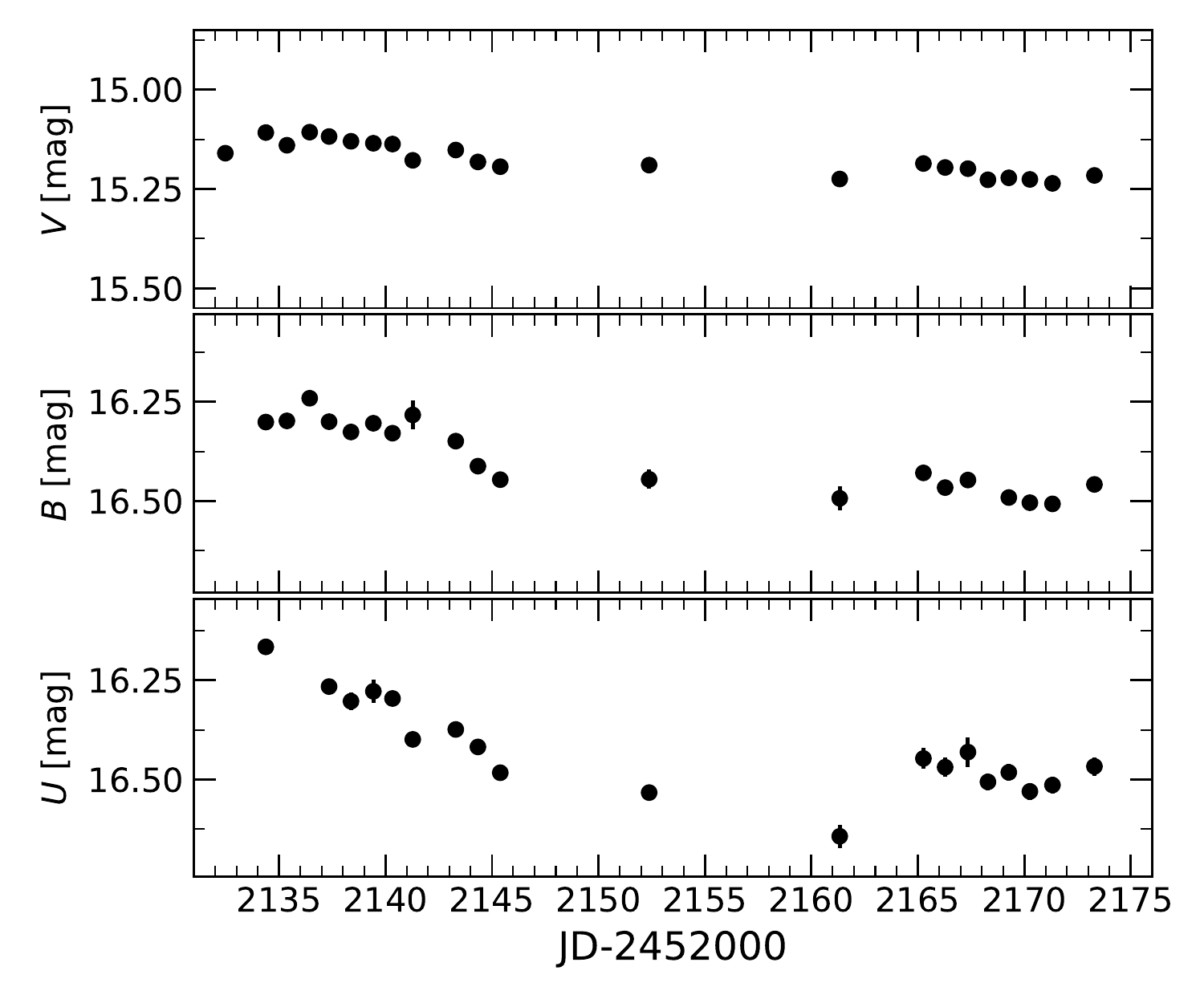}
\caption{$UBV$ photometry of V838 Mon taken with SAAO in the period from 2007 January 31 to March 13. 
All panels have the same y-axis extent of 0.7 mag for easier comparison of the variability.
The error bars are typically smaller than the  symbols.}
\label{F-saao}
\end{figure}

After the short eclipse-like event V838 Mon recovered to the pre-eclipse brightness, 
only to enter into a deeper minimum. During the first 1.5 months of the second decline our SAAO 
$UBVRI$ data were obtained (grey area in Fig.~\ref{F-lcour}). 
V838 Mon was observed on 22 nights, from 2007 January 31 to March 13.
Figure~\ref{F-saao} shows the SAAO $UBV$-band observations where the overall 
decline in brightness is visible.
Noteworthy is also the day-to-day variability of V838 Mon, which is most likely inherent to the system. 
Just as the overall decline is more pronounced in $UBV$, 
so is the daily variability larger in bluer bands, while the brightness is almost unchanged in $R$ and $I$.

Between March 2007 and the beginning of 2008 V838 Mon continued to fade, and it took  
more than two years to reach to the lowest brightness 
in $U$ band (see Fig.~\ref{F-lcour} and \citealt{2008ATel.1821....1G}). 
The deep fading affected mostly the  blue bands. 
The drop in brightness was around 5 magnitudes in $U$ and 3 magnitudes in $B$, 
while it was only about 0.5 magnitude in $V$ and almost unnoticed in $R$ and $I$. 
Our homogeneous set of NOT observations starts when the deep minimum was reached (around  2009). 
Since then the light curve has shown a slow but steady brightening in all bands.
A gradual brightening was 
detected before the 2006 eclipse-like event as well (see year 2004 in Fig.~\ref{F-lcour} 
and \citealt{2005MNRAS.358.1352C}).  
Variabilities with timescale
of about a month are noticeable, especially in the $B$ and $R$ bands, 
for which we have the best temporal coverage. 
This variability is more clearly seen in Fig.~\ref{F-lc2009B}. 
It displays very similar patterns in both the $B$ and $R$ bands.
This variability is further analysed in the next paragraph. 
Finally, the change in the colours (see Fig. 5 and GOR20) 
implies that V838 Mon has become bluer during the past decade, approaching a constant value since about 2018.

\begin{figure}[!t]
\centering
\includegraphics[width=9cm]{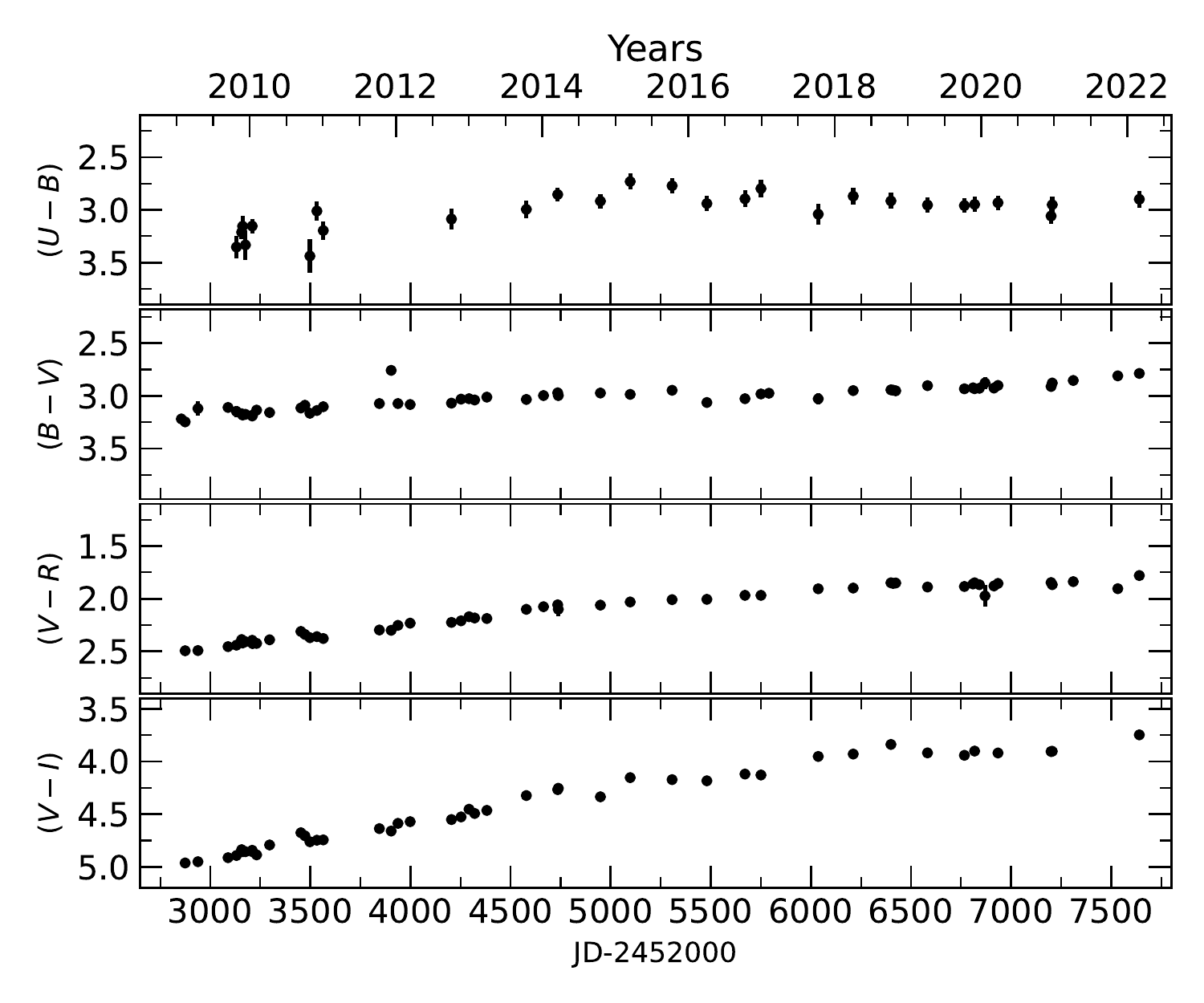}
\caption{Temporal evolution of the NOT colours since 2009. The y-axis in each panel spreads 
over 1.8 mag for easier comparison of the variability. }
\label{F-lcournot}
\end{figure}



\begin{figure}
\centering
\includegraphics[width=9cm]{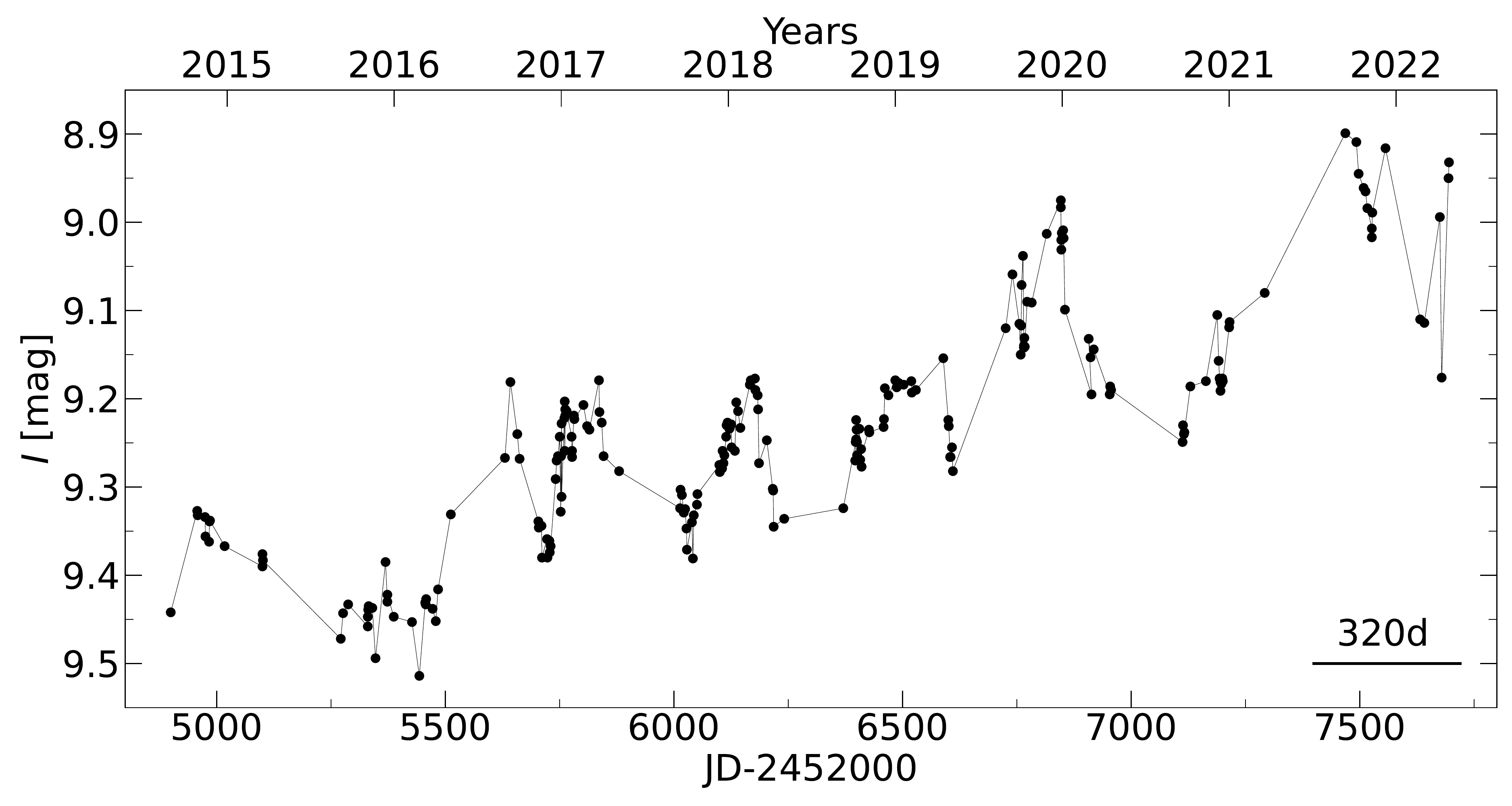}
\caption{$I$-band light curve created with the data from GOR20 and the following three years.
}
\label{F-I320}
\end{figure}

\textit{Period analyses. }GOR20 found in their $I$-band light curve, 
starting from the year 2017 (see also their Fig. 12), 
quasi-periodic variability with a possible period of about 320 days. 
The light curve variability during the past decade, with a similar timescale, is also evident   
in other photometric bands (see Fig.~\ref{F-lcour} and \ref{F-lc2009B}). 
We decided to check  the potential periodical nature of these fluctuations more thoroughly. 
Firstly, we recreated  \hbox{Fig. 12} from GOR20 together with their 
additional  data points for the past three years, 
which were collected after their publication and are freely available from their 
online table.\footnote{http://www.vgoranskij.net/v838mon.ne3} 
The resulting $I$-band light curve is presented in Fig.~\ref{F-I320}. 
It is clear that the 320-day semi-periodic variability does not continue after the year 2019. 
Secondly, to obtain more quantitative results, we combined our NOT data with the 
comparable GOR20 data in the  $BVRI$ bands and performed period analyses using a Lomb-Scargle periodogram 
(\citealt{1976Ap&SS..39..447L,1982ApJ...263..835S}), which is suitable for unevenly 
spaced data points. We note here that  we did not find a clear periodic variability. Therefore, 
we refrain from showing periodograms or folded light curves. However, we do mention 
indicative variations on  timescales of $\sim$100-300 days in all filters. In addition,  
a repeating period of $\sim$49 days in all bands was found.


\subsection{Spectral energy distribution}\label{S-sed}

The homogeneous set of photometric observations collected with   NOT provide us
a possibility to study changes in the spectral energy 
distribution (SED) of V838 Mon during the past decade when the system  experienced the deep minimum. 
In our analyses we take into account the recent ALMA observations
(\citealt{2021A&A...655A..32K}), which 
clearly show that both stars are surrounded with a distinct circumstellar 
dust cloud and that the  SG is not large enough to engulf the B3V star. 
The extensive SED analyses of the GOR20 concentrate mostly for the years before 
the current deep fading and cannot therefore be directly compared 
with our estimates.

In order to obtain the SED, the conversion from magnitudes to fluxes was done using 
zero magnitude fluxes  from \cite{2014MNRAS.444..392C}. 
For dereddening of the photometric data we followed earlier works of 
\cite{2005A&A...436.1009T}, \cite{2007A&A...474..585M}, 
\cite{2009ApJS..182...33K}, and \cite{2012A&A...548A..23T},
and adopted the standard extinction law $A_V = R_V E_{(B-V)}$ with a value of \hbox{$R_V = 3.1$}, 
the interstellar extinction curve of  \cite{1989ApJ...345..245C}, 
and a colour excess of V838 Mon of $E_{(B-V)} = 0.9$ mag from  \citet{2005A&A...436.1009T}.

\begin{figure}[!h]
\centering
\includegraphics[width=9cm]{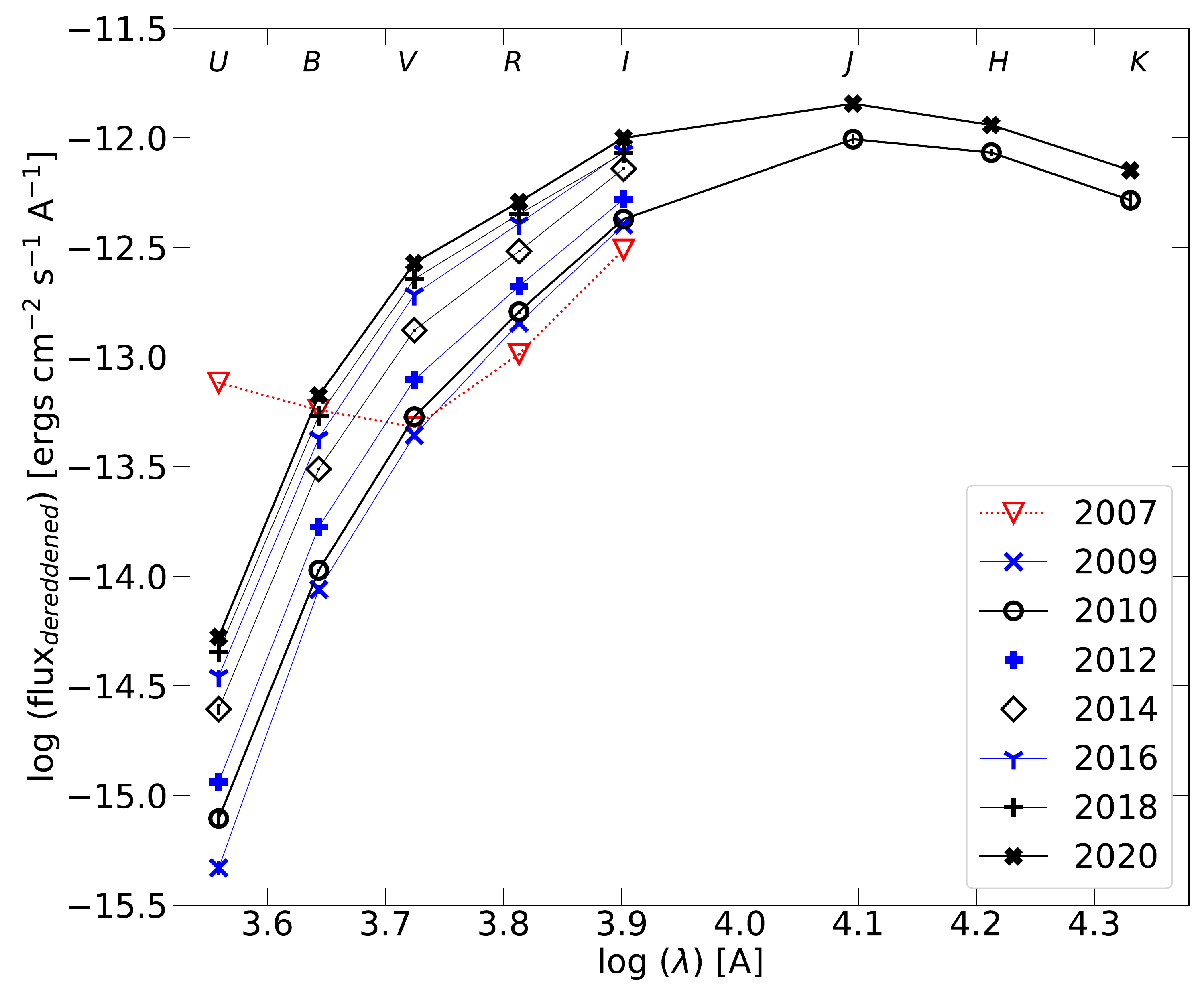}
\caption{SED of V838 Mon on selected years. 
Error bars are typically smaller than the symbols.
}
\label{F-sed}
\end{figure}

In Fig.~\ref{F-sed} we present 
the SED of V838 Mon from our NOT data 
at various selected dates between 2009 and 2020. For better visualisation, 
the photometric values are connected and the curves are shown alternating in black and blue. 
For comparison we also included in red the SED as it appeared in 
early 2007 after the short eclipse-like event based on our SAAO data.

It is obvious that in 2007 the blue component was still contributing
significantly to the SED, in particular to the $U$ and $B$ band,
whereas the red was dominated by the cool SG. During the
current deep minimum, the B3V star seems to be completely occulted. 
However, we note an overall increase in brightness (also shown in SED analyses of GOR20 using 
$UBVRI$ data), with more pronounced changes in
the bluer bands compared to the almost static behaviour in the infrared
($JHK_{BB}$) bands. This rise in fluxes might be caused either by a change
in the properties of the cool SG and its environment, by a steady
decrease in the absorption along the line of sight towards the B3V
companion, or  by a combination of the two, which we believe is most plausible (see
discussion in Sect.~\ref{S-disc}).

The many uncertainties and unknowns for such a scenario, especially the
unknown values of circumstellar reddening along the lines of sight towards
the two objects (which is most likely very different, see Sect.~\ref{S-disc}),
prevent us from performing a detailed analysis of the temporal evolution
of the SED,  but some simple estimations can be done. For example,
considering that the $U$-band flux in 2007 is entirely from the hot
companion, we can fit it with a model SED \citep[e.g.
ATLAS9,][]{2011MNRAS.413.1515H} utilising the parameters of a B3V star
($T_{\rm eff} \simeq 18\,000$\,K, $\log g \sim 4$) to obtain an estimate
of the stellar radius ($R_{\rm B3V} \sim 3.3$\,R$_{\odot}$) and luminosity
($L_{\rm B3V} \sim 10^{3}$\,L$_{\odot}$) for a distance of 5.9\,kpc
\citep{2020A&A...638A..17O}, in agreement with a main sequence star of
about 6\,M$_{\odot}$. On the other hand, if we assume that the $JHK$
fluxes in 2020 are purely from the red supergiant (RSG) with a proposed temperature of
$3500$\,K \citep{2021A&A...655A..32K} and $\log g = 0$, the SG's
current radius, resulting from fitting the near-IR fluxes with a
corresponding SED model, would be of the order of $450$\,R$_{\odot}$ in
agreement with a strongly inflated object. However, the reliability of
these values requires the (unlikely) case that both stars experience no
circumstellar extinction.

A further simple test can be done regarding the circumstellar extinction
towards the B3V companion. Using the same model SED as before, a
circumstellar dust composition similar to the interstellar composition, and
assuming that the $U$-band flux in 2010 and 2020 corresponds purely to the
hot companion, a minimum circumstellar extinction of $A_V= 2.7$\,mag and
$1.55$\,mag, respectively, is needed to suppress the star's blue emission.
Because  the contribution of the
cool component cannot be ignored for these measured low $U$ flux values, these are only a very crude lower limits
to the real extinction, but they might point towards a possible slight
recovery of the hot companion.


\subsection{Spectral analyses}\label{S-sp}

\subsubsection{Low-resolution spectra}\label{S-lr}

\begin{figure*}
\centering
\includegraphics[width=18cm]{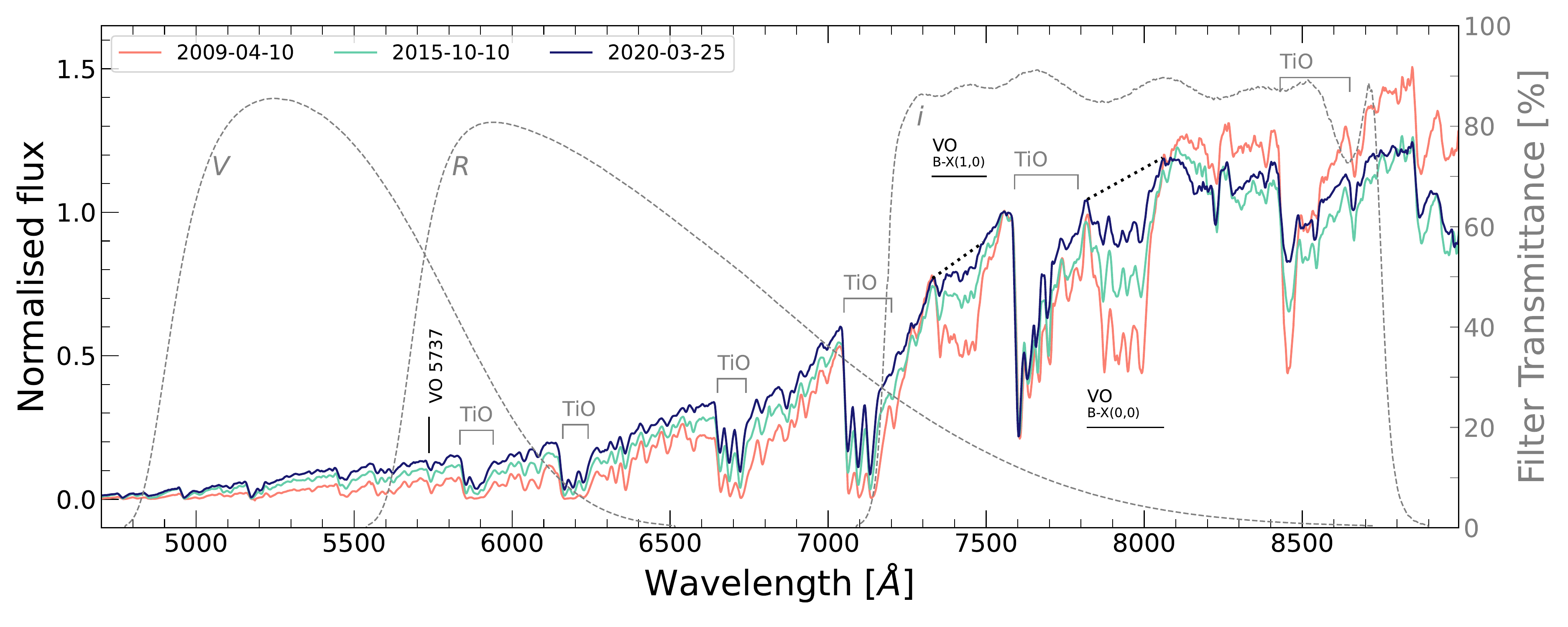}
\caption{Example low-resolution spectra for three selected epochs with the 
most prominent molecular band features identified. Normalisation of the calibrated spectra is 
performed at  $\lambda = 7560$\,\AA. 
Black dotted lines exemplary mark the pseudo-continuum for measuring the 
\hbox{(pseudo-)equivalent} widths of the VO band structures B-X(1,0) and B-X(0,0) 
in the 2020 spectrum. The NOT transmission curves of the ALFOSC $V$, $R$, and $i$ 
filters are shown with grey dashed lines.
}
\label{F-g4}
\end{figure*}

The spectral evolution of V838 Mon is depicted in Fig.~\ref{F-g4}. At each epoch intense molecular 
absorption bands dominate the spectral appearance. Most pronounced are the bands of vanadium 
oxide (VO) and titanium oxide (TiO). A multitude of absorption bands from these two molecules 
is typically seen in the atmospheres of stars with spectral type late-M 
\citep{1999ApJ...519..802K, 2009ApJS..182...33K, 2017ars..book.....L} and has been reported to 
dominate the spectrum of V838\,Mon following its outburst in 2002 
(e.g. \citealt{2009ApJS..182...33K}, GOR20).

For easier comparison, the calibrated spectra were normalised at 7560\,\AA, which we 
consider to be a line-free region (i.e. a continuum region). In this way, the intensities match at a 
single wavelength, but the change in the shape of the continuum remains conserved. The 
individual spectra shown in Fig.~\ref{F-g4} were selected to best represent the changes over the 
entire observing period, and the remaining spectra (not shown)  fall   between the limiting 
cases shown by the red and blue curves. 
Inspection of the spectra reveals a remarkable 
secular trend both in the spectrophotometric gradient and in the depth of molecular absorption 
features.

The low resolution of our spectra limits the usability of the molecular bands, which 
otherwise provide a superb tool for characterising  their formation region, 
as  demonstrated by \cite{2009ApJS..182...33K} 
based on high-resolution spectra. However,  we identified
a few VO bands (highlighted in Fig.~\ref{F-g4}) that appear to be   
measurable and the least contaminated by neighbouring 
absorption features in the complex spectrum of V838 Mon.

\begin{figure}
\centering
\includegraphics[width=9cm]{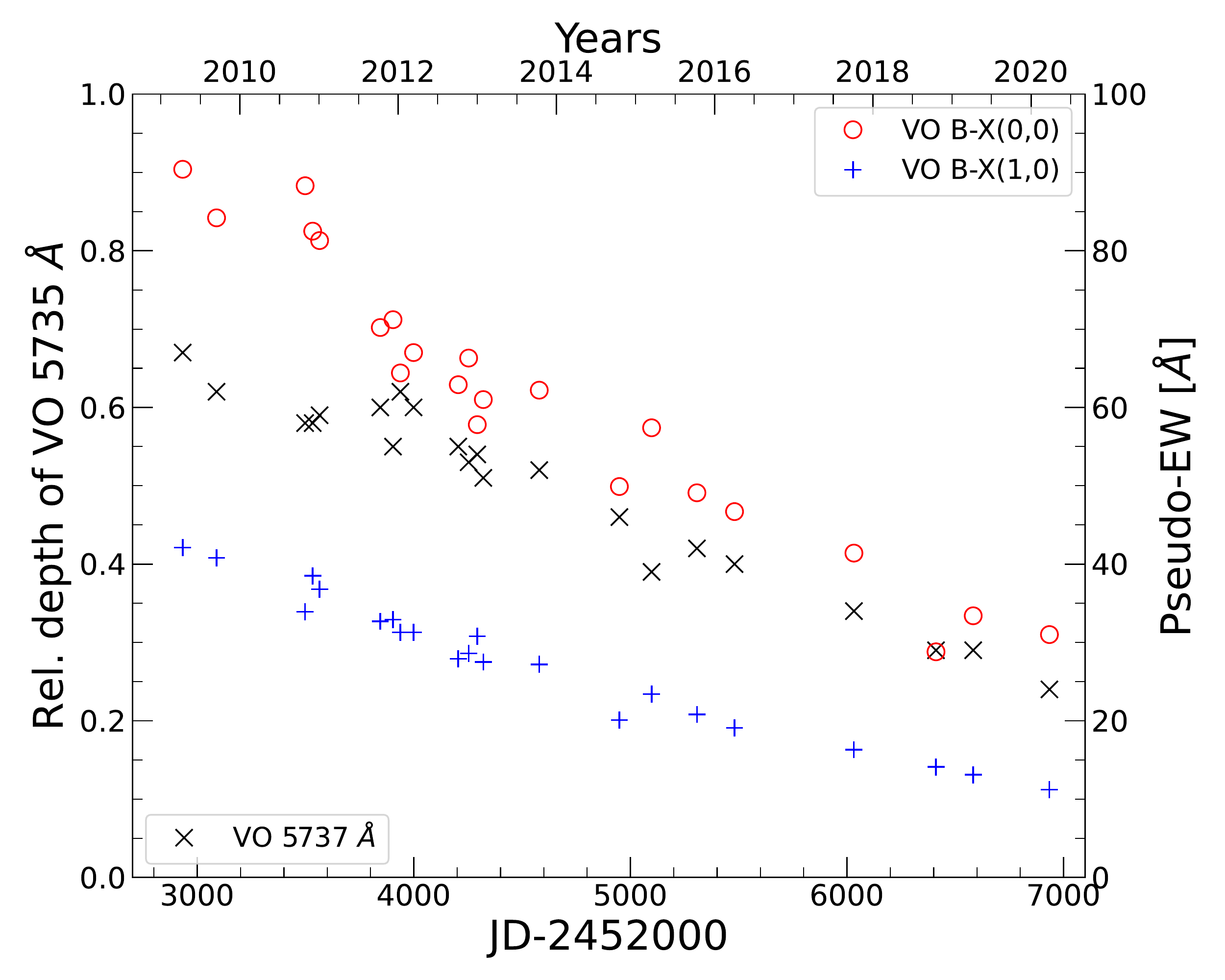}
\caption{ Temporal evolution of the relative strength (black crosses, left 
scale) of the VO band head C-X(0,0) R$_{4}$ absorption feature and the B-X(1,0) (blue plus 
signs) and B-X(0,0) (red circles) equivalent width measurements (right scale).
}
\label{F-vo_evol}
\end{figure}

For all observing epochs, we  measured the depth of the VO band head C-X(0,0) R$_{4}$  
($\lambda_{\rm lab} = 5736.703$\,\AA) relative to the adjacent continuum.
In addition, we measured the pseudo-equivalent 
widths of the VO band structures B-X(1,0) and B-X(0,0) within the wavelength limits 7330 to 
7500\,\AA \ and 7820 to 8060\,\AA, respectively, as shown in Fig.~\ref{F-g4}. The measurements of 
the absorption depth and of the equivalent widths throughout our entire spectral series are 
listed in Table~\ref{T-vo}, and their temporal evolution is shown in Fig.~\ref{F-vo_evol}. A clear, continuous 
decrease in the strength of the band head and in the equivalent widths by more than a factor 
of two can be seen during the 11-year observing interval.

\begin{table}
\caption{Measured relative depths of the VO band head C-X(0,0) R$_{4}$  
($\lambda_{\rm lab} = 5736.703$\,\AA) and pseudo-equivalent widths of the VO band 
structures B-X(1,0) and B-X(0,0) from our low-resolution spectra together with the 
respective $(V-I)$ colours. 
The entire table is available in electronic form at the CDS. Here only the first
five lines are presented for guidance.}              
\label{T-vo}      
\centering                                      
\begin{tabular}{c c |c c c}          
\hline\hline                        

              &         &          &VO  & \\
  JD-2452000  & $(V-I)$ & \textit{rel. depth} & B-X(1,0)& B-X(0,0)  \\ 
   $[$days$]$     & [mag] & $5737\AA$ & $[\AA]$ & $[\AA]$  \\
\hline                           

2932.36&4.95  & 0.67 & 42.1 & 90.4  \\
3088.74&4.91  & 0.62 & 40.8 & 84.2  \\
3497.72&4.76  & 0.58 & 33.9 & 88.3  \\
3532.61&4.75  & 0.58 & 38.5 & 82.5  \\
3564.49&4.75  & 0.59 & 36.8 & 81.3  \\

\hline                                             
\end{tabular}
\end{table}


\begin{figure}
\centering
\includegraphics[width=9cm]{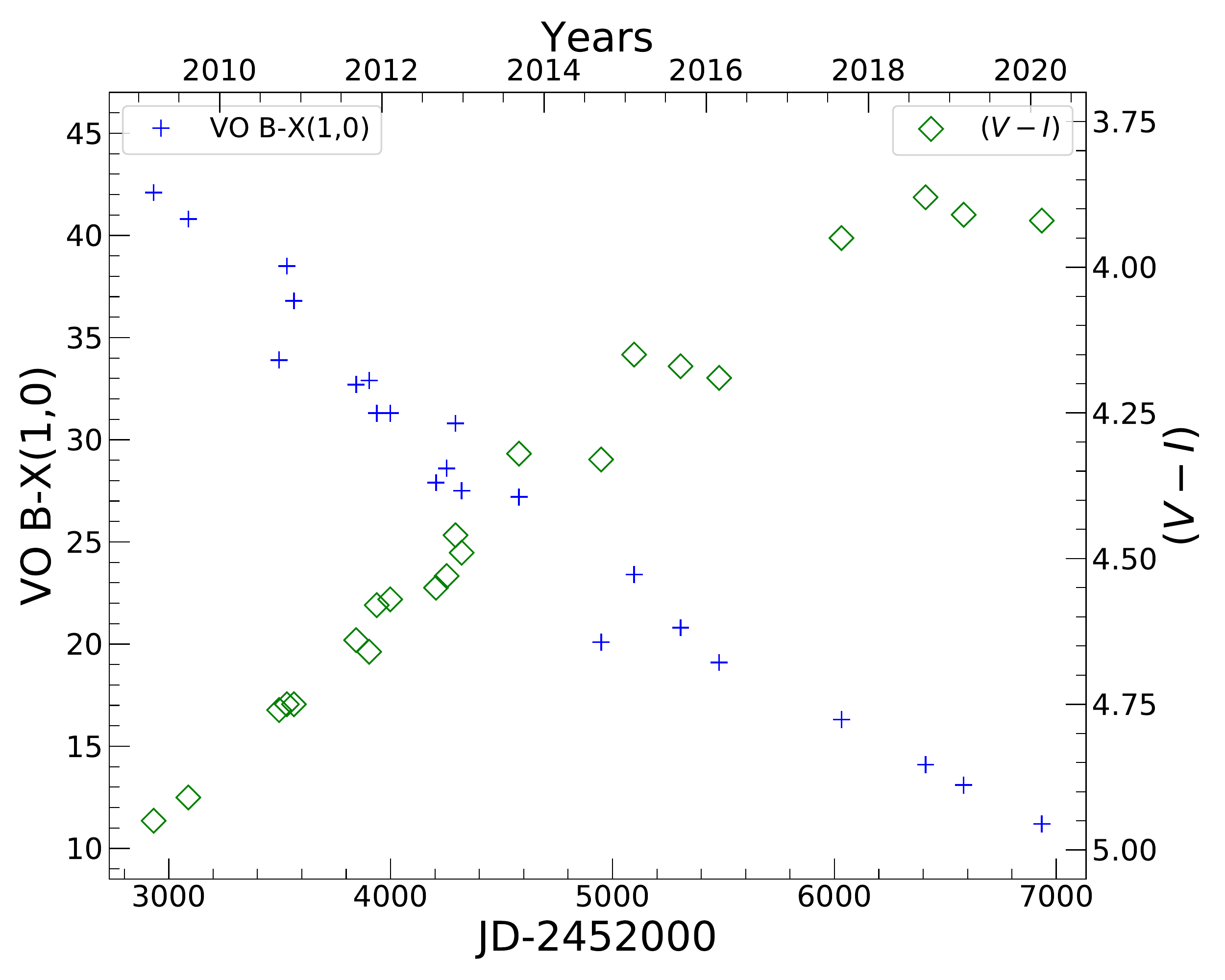}
\caption{Comparison of the temporal evolution of the VO B-X(1,0) (blue plus 
signs, left scale) equivalent width measurements and the $(V-I)$ colours (green diamonds, 
right scale).
}
\label{F-vo_vi}
\end{figure}

Our chosen normalisation also reveals a decrease in the spectrophotometric gradient, 
meaning that the continuum was redder in 2009 and has continuously become bluer since then 
(see Fig.~\ref{F-g4}), which is best described by the $(V-I)$ colour measurements  
in Table~\ref{T-vo} and depicted in Fig.~\ref{F-vo_vi}. 
The possible origin of this trend is  discussed in Sect.~\ref{S-disc}

\subsubsection{Medium-resolution spectra}\label{S-hr}

\begin{figure*}
\centering
\includegraphics[width=18cm]{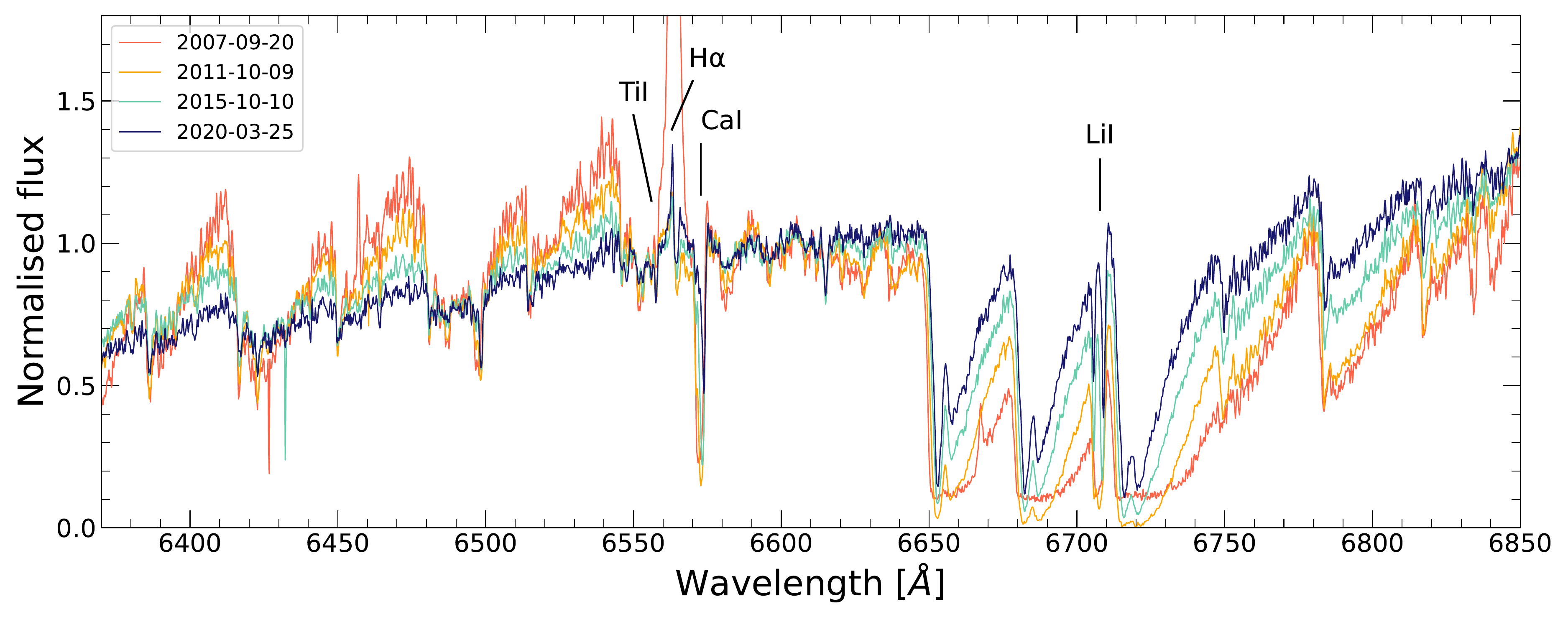}
\caption{Example medium-resolution spectra for three selected dates with the 
most prominent atomic features identified. As a comparison, a spectrum from before the deep minimum (2007) 
is plotted in red. Normalisation of the calibrated spectra is 
performed at $\lambda = 6577$\,\AA.}
\label{F-g17}
\end{figure*}

Our medium-resolution spectra spread from 6360 to 6860\,\AA. 
This wavelength range does not include the 7560 Å used for normalisation of the low-resolution spectra.
Therefore, the normalisation of the calibrated spectra had 
to be done at a different position, for which we selected a central, presumably line-free 
region around 6577\,\AA. For demonstration purpose of the linear temporal behaviour of the 
spectral features, we show in Fig.~\ref{F-g17} three example spectra spreading from 2011 to 2020, 
together with an archival spectrum from 2007 before the disappearance of the hot binary component in 2008.
As for the low-resolution cases, the spectra are dominated by the molecular absorption bands of 
TiO, and the weakening of these bands with time is clearly seen. In addition to the molecular 
bands, the high-resolution spectra also display prominent atomic lines. These are the two 
resonance lines (Ca\,{\sc i} 6572.78\,\AA, Li\,{\sc i} 6707.83\,\AA) and H$\alpha$. They are  
identifed in Fig.~\ref{F-g17} along with the weak Ti\,{\sc i} 6556.06\,\AA \ line. 
This latter line serves as an example of relatively stable features in the spectra of V838~Mon.

\begin{figure}
\centering
\includegraphics[width=9cm]{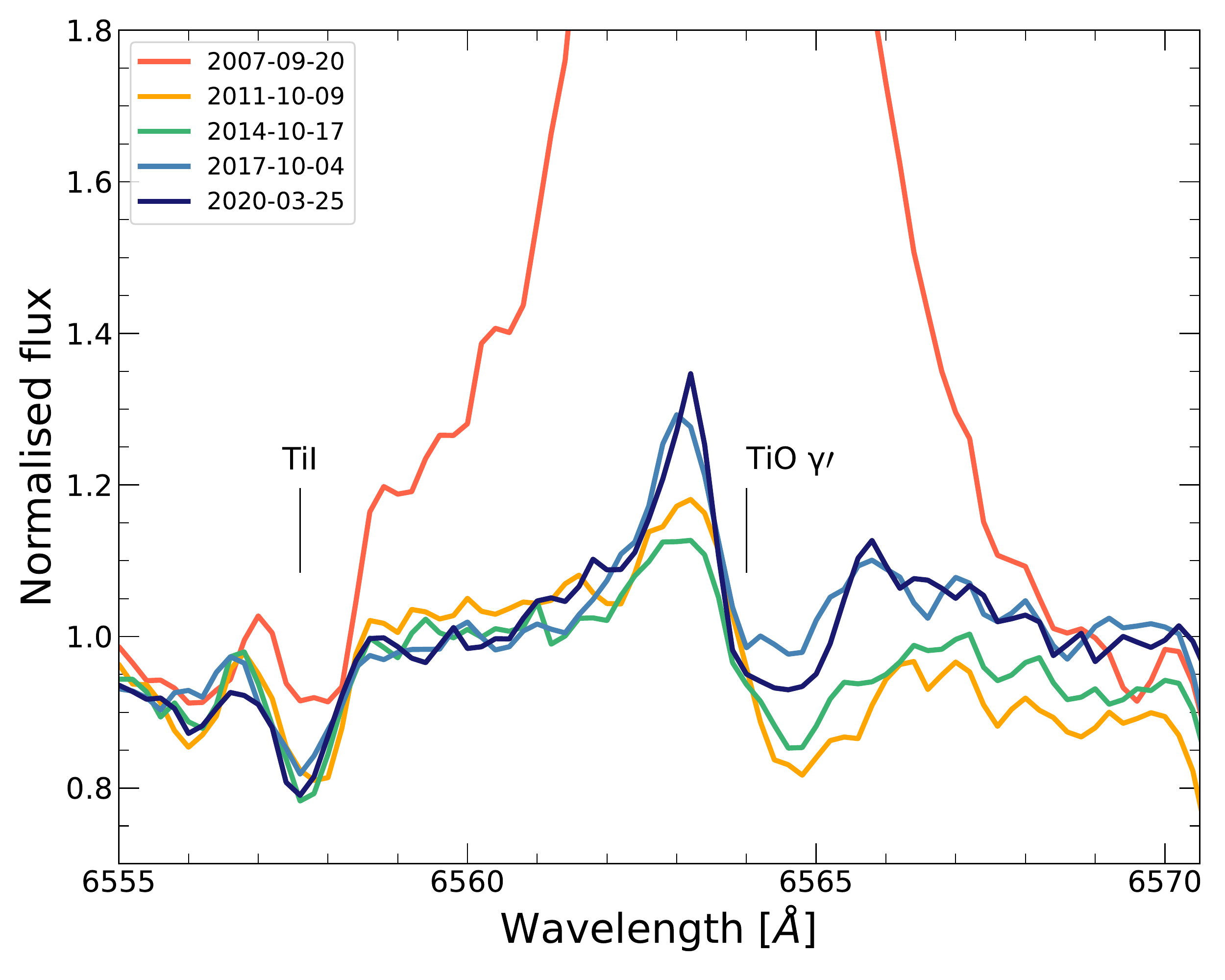}
\caption{Temporal evolution of the H$\alpha$ line showing a very slowly 
proceeding recovery trend of the previously intense emission.}
\label{F-g17ha}
\end{figure}

The region around H$\alpha$ is depicted in more detail and for more epochs in Fig.~\ref{F-g17ha}. 
The H$\alpha$ emission in 2007 was very strong, 
whereas it was only a minor feature between 2011 and 2014. Only thereafter does it seem that 
the emission component has started a (very slow) recovery.

When interpreting the shape and intensity of the H$\alpha$ line one should be aware that 
the profile might be contaminated with the band head of the TiO $\gamma^{\prime}$ (0-1) 
band arising at $\sim 6564$\,\AA \ \citep[see identification in][their 
Fig.~1]{2011A&A...532A.138T}. The superposition of the TiO absorption band with the red wing 
of the H$\alpha$ emission makes the H$\alpha$ profile appear like an inverse P~Cygni line.
For clarity we   indicate the position of the TiO 6564\,\AA \ band head in 
Fig.~\ref{F-g17ha}.

\begin{figure}
\centering
\includegraphics[width=8.5cm]{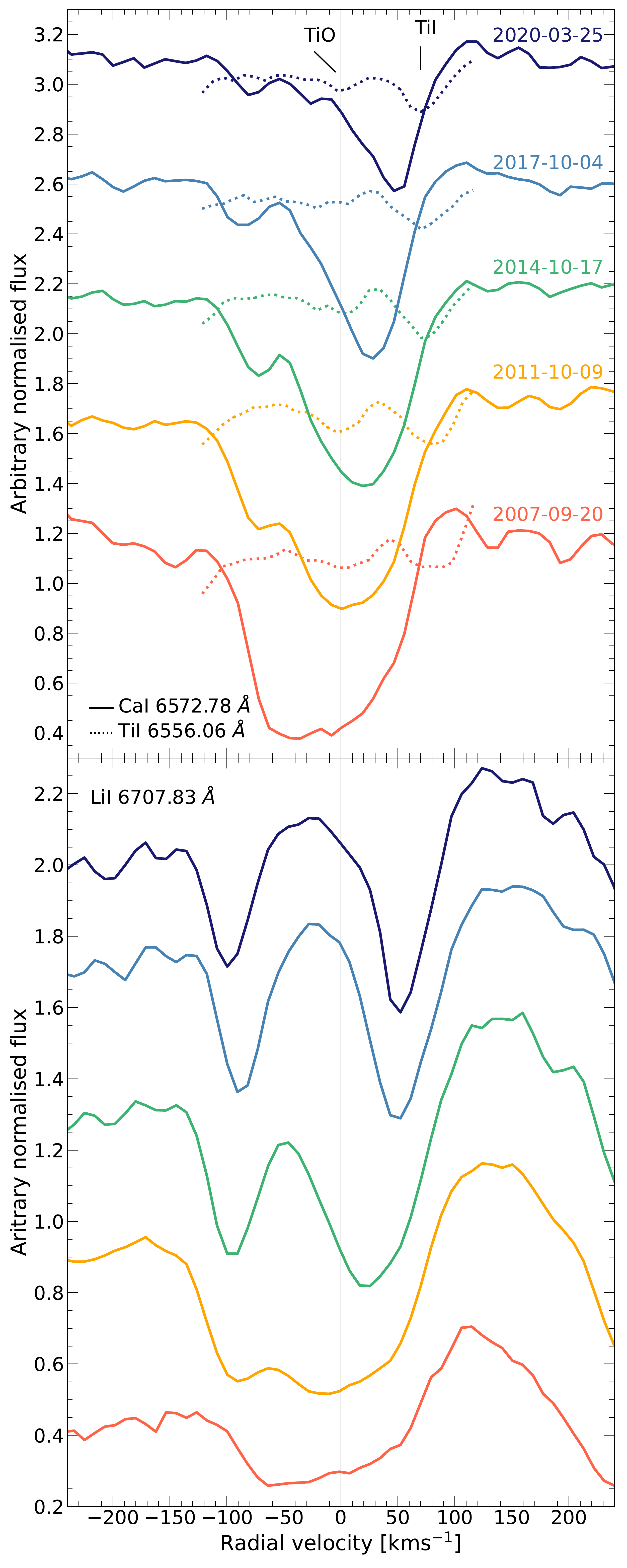}
\caption{Comparison of the profiles of Ca\,{\sc i} 6572.78\,\AA \ (solid 
lines) and Ti\,{\sc i} 6556.06\,\AA \ (dotted lines) for selected dates, highlighting the 
stability in radial velocity of the Ti\,{\sc i} 6556.06\,\AA \ line \textit{(top panel)}.
Temporal evolution of the Li\,{\sc i} 6707.83\,\AA \ line showing a drastic change from a P Cygni-type
profile  in 2007 to a double-absorption profile and a subsequent weakening of the 
absorption components (\textit{bottom panel}). The colour-coding is the same for both panels. 
To guide the eye, the position of zero radial velocity is given by the grey vertical line. 
}
\label{F-g17cati}
\end{figure}

Turning to the profiles of the resonance lines of Ca\,{\sc i} and Li\,{\sc i}, their temporal 
variation is depicted in radial velocity scale in the top and bottom panels of 
Fig.~\ref{F-g17cati}, 
respectively, for five selected dates. In 2007 both lines displayed a P Cygni-type profile 
with a shallow red-shifted emission component and a single, broad blue-shifted absorption 
component. The emission component is more easily seen for Ca\,{\sc i,} whereas the Li\,{\sc i}
profile shape is influenced by the underlying wing of the adjacent molecular absorption bands.
This P Cygni-type profile has changed significantly during our monitoring 
period. A double-absorption has emerged with a broader and more intense red 
component. This is more clearly seen in Li\,{\sc i}, and is only apparent in the 2014 spectrum of 
Ca\,{\sc i}. This red absorption component moves further redwards with time, and a 
general weakening of the whole absorption is seen in parallel with the weakening of the 
molecular absorption bands.

Finally, we compare in Fig.~\ref{F-g17cati} the temporal evolution of the 
Ca\,{\sc i} 6572.78\,\AA\ line 
profile with that of the weak Ti\,{\sc i} 6556.06\,\AA\ line. Both lines are plotted in 
radial velocity scale for several selected dates. Interestingly, Ti\,{\sc i} appears rather 
stable in both its radial velocity (around 75 km\,s$^{-1}$) and absorption strength. 
Moreover, its radial velocity nearly coincides with the redmost wing of the Ca\,{\sc i} 
absorption. 

Possible reasons for the spectral variability seen during the past 13 years are
  discussed in Sect.~\ref{S-disc}.

\section{Discussion}\label{S-disc}

Our photometric and spectroscopic monitoring of V838~Mon over more than a decade reveal 
\begin{itemize}
\item a steady brightening in all photometric bands, particularly in the optical (Fig.~\ref{F-sed}), 
\item a continuous weakening of the molecular absorption bands (Fig.~\ref{F-vo_evol}), 
\item a slight strengthening of the H$\alpha$ emission (Fig.~\ref{F-g17ha}), and
\item significant changes in the profile shapes and radial velocities of the resonance lines 
Ca\,{\sc i} 6572.78\,\AA \ and Li\,{\sc i} 6707.83\,\AA \ along with
a presumably stable radial velocity of Ti\,{\sc i} 6556.06\,\AA \, (Fig.~\ref{F-g17cati}).
\end{itemize}
Possible causes for these trends are   discussed in the following in the context of the current 
knowledge about the stellar and circumstellar properties of V838~Mon.

Since the eruption in 2002, V838~Mon appears as a cool SG star embedded in a large-scale, 
dense envelope of expanding material \citep{2011A&A...529A..48K, 2016A&A...596A..96E} and with an 
ongoing stellar mass-loss \citep{2009ApJS..182...33K}. Infrared and radio interferometric measurements revealed 
that this cool SG is surrounded by a disc-like dusty envelope \citep{2014A&A...569L...3C, 
2020A&A...638A..17O, 2021A&A...655A..32K, 2021A&A...655A.100M}. 
The dust has most likely condensed from the material 
ejected during the 2002 outburst \citep{2008ApJ...683L.171W} and causes intense mid- to far-infrared 
excess emission \citep{2021AJ....162..183W}. The ejected material, and in particular the extended 
dusty envelope seems to be responsible for the disappearance of the hot companion from both the 
spectra and the SED. 
However, in addition to the large-scale dusty envelope, the B3V companion 
seems to be embedded within a thick dusty, possibly spherical envelope on small scales 
\citep[see Fig.~6 in][]{2021A&A...655A.100M}. This envelope might represent material 
being accreted by the hot component.

Looking only at the photometric brightening (Fig.~\ref{F-sed}) and the clear decline in the colour $(V-I)$ 
(Fig.~\ref{F-vo_vi}) one could speculate that the observed change in SED suggests a reduction in circumstellar 
extinction due to the expansion, and hence dilution of the ejected 
material, and dilution of the spherical dust envelope around the B3V component due to 
ongoing accretion of material that is not further replenished 
and/or due to a reduced 
mass-loss from the SG star. The expanding molecular environment has been identified as 
contributing significantly to the observed molecular absorption \citep{2009ApJS..182...33K, 
2011A&A...532A.138T}. The continuous weakening of the molecular absorption bands could thus also be 
interpreted as being due to less dense absorbing material surrounding the star, and recent investigations
by \citet{2021MNRAS.508.5757D} have shown that the mass-loss of cool SGs significantly 
alters the strength of the molecular absorption bands. The circumstellar dust 
traced by interferometric observations seems not to be along the line of sight towards the cool 
SG, but confined to an elongated disc-like structure that is seen under a moderate inclination 
angle \citep{2014A&A...569L...3C, 2021A&A...655A..32K, 2021A&A...655A.100M}. 
In contrast, the wind of the SG might 
contain a dusty component, but in that case it was found 
to be semi-transparent (with an optical depth of 1.44 at 0.55\,$\mu$m, \citealt{2021AJ....162..183W})
or even optically thin, and hence plays no 
significant role for a possible circumstellar extinction \citep{2021A&A...655A..32K}. An alternative 
(or additional) explanation for the steady changes seen in both photometry and spectroscopy might
therefore be related to a change (increase) in effective temperature, especially because the 
molecular absorption bands are considered as important temperature indicators in cool SGs 
\citep[e.g.][]{2017ars..book.....L}.

From the spectroscopic point of view, V838~Mon appeared as an L-type star after its outburst in 
2002 \citep{2003MNRAS.343.1054E}, and the star had been assigned a SG state based on the 
strength of the molecular absorption bands in the optical \citep{2003MNRAS.343.1054E} and 
near-infrared spectra \citep{2007A&A...467..269G}. The low resolution of our spectra 
hampers a more detailed analysis and proper fitting with atmospheric 
models. Moreover, as has been shown by \citet{2013ApJ...767....3D}, the temperatures 
obtained from TiO band modelling turn out to be systematically lower than the temperatures 
derived from fitting the SED in the near-IR, and these authors consider the SED method to be 
the more reliable one. Nevertheless, our measurements of the depth and equivalent widths 
of some of the VO band features (see Fig.~\ref{F-vo_evol}) point towards changes in the spectral appearance that 
resemble an increase in temperature, at least in a qualitative way.  

\begin{figure}
\centering
\includegraphics[width=9cm]{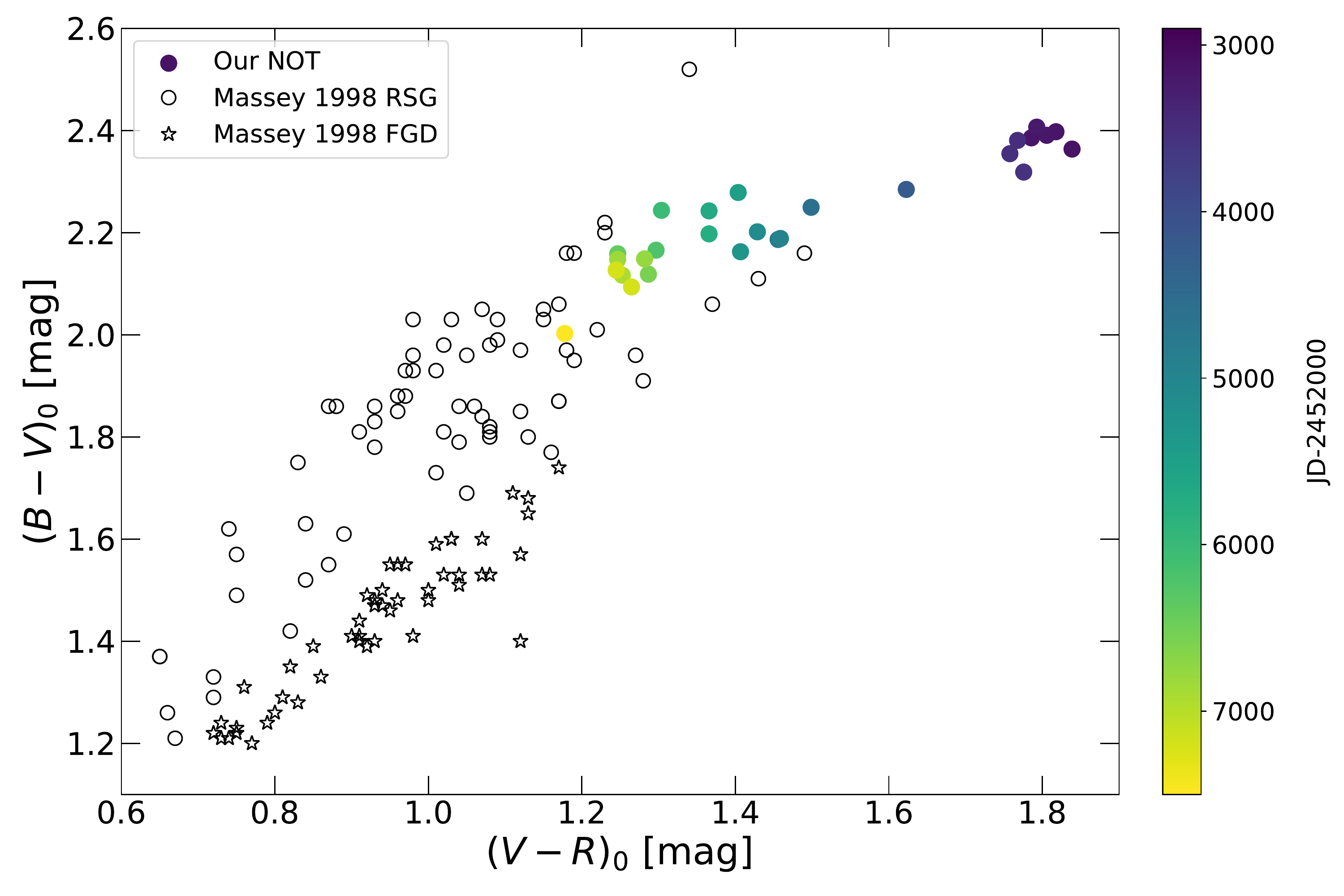}
\caption{Our dereddened colours plotted against the red supergiants (RSG) and foreground stars (FGD) 
from \citet{1998ApJ...501..153M}. The colour bar corresponds to our data at various epochs. 
}
\label{F-massey}
\end{figure}

A solid check for low surface gravity objects has been established by \citet{1998ApJ...501..153M} 
based on the displacement of SGs in the $(B-V)_{0}$ versus $(V-R)_{0}$ colour-colour diagram 
from dwarf stars with equal temperatures.  
In Fig.~\ref{F-massey} we display this colour-colour diagram for the
sample of red SGs (black circles) along with foreground (dwarf) stars (black stars), 
both taken from \citet[][their Table~2]{1998ApJ...501..153M}. 
Their data include photometrically classified as well as spectroscopically 
confirmed RSGs in three galaxies in 
the Local Group: NGC 6822, M33, and M31 (see also their Fig.~10).
Our dereddened measurements of V838~Mon are included in 
this diagram, and their colour-coding refers to the observing epochs. We note a clear temporal 
evolution of V838~Mon in this diagram, starting in the upper right corner 
(far beyond even the coolest normal red supergiants) and 
gradually approaching the region of the coolest RSG. 
We also note a clear separation between the positions of 
V838~Mon and the dwarf stars in this diagram. Taken together, these two facts confirm that the star is 
a strongly inflated low-gravity object whose effective temperature has seemingly increased during the 
past 13 years that were covered by our observations.

With the knowledge that V838~Mon resembles the coolest RSG, we utilise the power-law relation between the intrinsic colour $(V-K)_{0}$ and the effective temperature that was derived by 
\citet{2005ApJ...628..973L} for classical RSG. This relation can only be applied to our 
data from 2010 and 2020 due to the lack of near-infrared photometry for all other epochs. 
As mentioned in Sect.~\ref{S-irdata}, our $K_{s}$ magnitudes were transformed into the 
needed \citet{1988PASP..100.1134B} photometric system.
Our values for $(V-K_{BB}$)$_{0}$ of $7.27\pm 0.11$\,mag and $5.86\pm 0.06$\,mag 
deliver effective temperature values of 
$3260\pm 50$\,K and $3390\pm 50$\,K in the years 2010 and 
2020, respectively, and thus point to a temperature increase of $\sim 130$\,K within ten 
years.
We can further exploit our $(V-K_{BB})_0$ colour and estimate the temperature using the 
empirical colour versus temperature calibration of RSG by \citet{2011ApJS..193....1W}.
From our dereddened $(V-K_{BB})$ colours  
we obtain the $T_{\text{eff}}$ to be 3170$\pm$50 K and 3390$\pm$50 K for 2010 and 2020, respectively. 
For this we adopted the solar metallicity (Z=0.02) (suitable assumption 
for V838 Mon according to \citealt{2004A&A...416.1107K}) and log(g)=0  
that were  used 
for V838 Mon by \cite{2009ApJS..182...33K} and \cite{2011A&A...532A.138T}.
These temperature estimates are in agreement 
with our temperature calculations using the power-law relation by \citet{2005ApJ...628..973L}. 
Therefore, we conclude that the change in
$(V-K_{BB}$)$_{0}$ colour between 2010 and 2020 indicates a change in temperature
of the giant of $130-220$ K. 
Both of our temperature estimates using the $(V-K_{BB})_0$ colour
compare well with those of 
\citet{2021A&A...655A..32K}, who obtained values of 3300\,K and 3500\,K from modelling of their spectra
taken in 2012 and 2020, respectively.
It is noteworthy that  these temperature values are considerably higher 
than what was   obtained from Planckian fits found 
in the literature (e.g. \citealt{2015AJ....149...17L}), and emphasises the poor representation of cool
SG stars' SEDs with black-body functions. 
On the other hand, these temperature values are also lower than  is typically seen for 
classical Galactic RSGs, which seem to have a lower temperature limit of around 
$3500-3600$\,K \citep[see e.g.][]{2021MNRAS.502.4210T}. This peculiar behaviour of V838 Mon suggests 
that its (apparent) photosphere may be more inflated and cooler, similar to very cool (lower mass) giants.
 
The rise in blue continuum along with the weakening of the molecular absorption bands suggest
a trend for increasing effective temperature of the cool SG star; however,  we should keep in mind that 
the B3V companion, which we have so far neglected in our discussion, also contributes to the total 
continuum. 
The disappearance of this companion is interpreted as being due to the engulfment of the star by 
the extended dusty envelope and the formation of
a compact dust shell embedding the secondary on much smaller scales \citep{2021A&A...655A.100M}. 
This compact shell was created by accretion of the dust from the reservoir provided by 
the matter ejected during the 2002 outburst, and is further fed by wind material released from the RSG.
Hence, the rise in mainly the $U$- and $B$-band fluxes during the past decade 
could also be interpreted as a slow but continuous recovery of the hot companion (see  discussion in Sect.~\ref{S-sed}). 
Support for this  
interpretation is provided by the fact that the flux in the mid-infrared ($> 10\,\mu$m) considerably 
decreased between 2008 and 2019 \citep{2021AJ....162..183W}, indicating that a significant amount of 
warm dust has disappeared either due to expansion and cooling or due to evaporation by the
radiation from the hot companion. 
Furthermore, the steadily decreasing wind of the RSG (as we show below)
cannot replenish and maintain the dust reservoir from which the material
is accreted to the secondary, resulting  in a dilution of the
circumstellar matter around the B3V companion. 
In either case, the medium hiding the hot companion has started to 
become more transparent. In this context we note that the H$\alpha$ emission, 
which originates from the hot companion, gradually increased during the past decade (Fig.~\ref{F-g17ha}). 
Compared to the spectra presented by \citet[][their Fig.~2]{2015AJ....149...17L}, in which the 
emission component of H$\alpha$ during the observing period between 2010 and 2012 reached values of  
$1.1-1.2$ times the continuum, our latest observations reveal an H$\alpha$ emission up to 1.3 times 
the continuum.\footnote{A proper estimate of the emission flux in H$\alpha$ requires detailed modelling 
of the superimposed TiO $\gamma^{\prime}$-band absorption (see Fig.~\ref{F-g17ha}), 
which is beyond the scope of the present investigation.} 
This obvious increase in H$\alpha$ emission, also reported by GOR20 
based on their data from 2012 to 2018, supports the suggestion of 
a gradual reappearance of the hot companion. 

Based on the resemblance of the V838 Mon spectrum with spectra
of RSG we compare the characteristics of its light curve
with corresponding features amongst ordinary RSGs 
(cf. \citealt[Chapter 7]{2017ars..book.....L} and \citealt{2006MNRAS.372.1721K}). 
We note the similarity of the behaviour of V838 Mon to
the sub-class of irregularly variable RSGs, both in the values of
amplitudes (approximately $\pm$ 0.2 mag) and timescales (hundreds of days).
The 49-day cycle may not strictly follow this trend.
The milder amplitude in the I band is similar to RSGs as well.
We note that the possible reasons for the photometric variability
also seem  be common with RSGs: pulsations, convection at the surface,
mass-loss, and dust production variability.

Next we turn to the temporal evolution in the resonance line profiles 
shown in Fig.~\ref{F-g17cati}. In 2007   the Ca\,{\sc i} and Li\,{\sc i} lines both displayed pure P Cygni profiles, while a splitting
into two absorption components becomes evident for our data from 2011 on. According to the analysis 
by \citet{2011A&A...532A.138T}, the bluer of these absorption components is formed in the matter 
ejected in 2002 and presumably represents a fast moving shell. A comparison of the strengths of 
the shell absorption components in both Ca\,{\sc i} and Li\,{\sc i} implies that the opacity in the 
Li line is higher than in the Ca line. Because the opacity drops with the drop in density of the 
expanding shell material, which for an assumed constant expansion velocity becomes proportional to 
$r^{-2}$ where $r$ is the distance from the star, the weakening of the less opaque shell component in 
Ca and its slightly smaller blue-shifted velocity compared to Li is evidence for a density and 
velocity gradient within the expanding shell. In contrast, the redder of the two absorption components
is similar in both lines and is formed most probably in the ongoing wind of V838~Mon. The existence
of such a wind is evident from the representative P Cygni-type profiles of the persistent K\,{\sc i} 
7698\,\AA \ resonance line presented in \citet[][their Fig.~6,]{2007ASPC..363...13M} and 
\citet[][their Fig.~7,]{2021A&A...655A..32K}   over the years 2002--2020. This potassium 
line is a good example to demonstrate that even at epochs when the Ca\,{\sc i} and Li\,{\sc i} 
resonance lines suggest a weakening of the wind (see Fig.~\ref{F-g17cati}), the high ground-level
population of K\,{\sc i} still results  in saturated wind profiles over all Doppler velocities
\citep[Fig.~7 in][]{2021A&A...655A..32K}. A similar saturation was present in Ca\,{\sc i} and 
Li\,{\sc i}   in 2007.

Finally, we turn our attention to the atomic absorption line Ti\,{\sc i} 6556.06 \AA, 
which is present in our medium-resolution spectra 
(Figs.~\ref{F-g17ha} and \ref{F-g17cati}) and has had a  
stable position between the years 2009 and 2020 with a radial velocity 
of \hbox{$+75\pm 2$\,km\,s$^{-1}$} and a full width at the continuum level of \hbox{85 km\,s$^{-1}$.}
When comparing the Ti\,{\sc i} profile 
in our spectra with profiles of Ti\,{\sc i} lines (6556.06 \AA\ line among them) 
in higher resolution spectra (see Fig. 3 panel c in \citealt{2011A&A...532A.138T}) 
it is clear that the line profiles are 
asymmetric, displaying a clear blue wing or blend. This asymmetry,
when fitting the entire profile in our significantly lower resolution data
with a single Gaussian, results in a systematic underestimation of the
line's radial velocity. Figure~3 of \cite{2011A&A...532A.138T} clearly shows  that the real radial velocity of the Ti\,{\sc i} line has a value
closer to +90 km\,s$^{-1}$. 
A similar narrow velocity 
component of $86-87$\,km\,s$^{-1}$ has been found in the lines of the CO fundamental and first 
overtone bands \citep{2007A&A...467..269G} and  in a number of atomic lines and high-excitation 
molecular bands \citep{2009A&A...503..899T}, and has been interpreted by these authors as infall. In 
particular the lines of V\,{\sc i} \citep[see][their Fig.~3]{2009A&A...503..899T} display only this 
red-most velocity component, just as our Ti\,{\sc i}  6556\,\AA \ line. Considering that in the cooler 
envelope (or wind) these atoms are completely bound in the molecules TiO and VO, the absorption lines 
of atomic Ti\,{\sc i} and V\,{\sc i} must arise in a hotter region in (or close to) the stellar 
photosphere. Support for this  interpretation comes from the fact that the excitation potential of 
the lower level of the Ti\,{\sc i}  6556\,\AA \ transition is 1.46 eV, whereas most of the other 
atomic lines observable in the optical spectrum of V838~Mon with wind-modified profiles have lower 
excitation potentials \citep{2011A&A...532A.138T}. In addition, \cite{2021A&A...655A..32K} 
have measured in the 2020 SALT high-resolution spectrum of V838 Mon the absorption components 
with heliocentric velocity of +85 km\,s$^{-1}$ in several lines from highly excited 
states. These pieces of evidence and
the persistence of this absorption line 
\citep[and red-most absorption component in many other lines, see Fig.~3 in][]{2009A&A...503..899T}
for more than 15 years speaks against an interpretation as matter infall or stellar contraction:  a contraction with a constant velocity of $16$\,km\,s$^{-1}$ over a time span of 15 years 
would mean that the material has travelled a distance of more than $10^{4}\,R_{\sun}$. 
\citet{2014A&A...569L...3C} reported on a possible contraction of the SG's radius by 
$\sim 40\%$ from about $1200\,R_{\sun}$ to about $750\,R_{\sun}$ within a ten-year period. Such a 
contraction causes a red-shifted velocity of just $\sim 1$\,km\,s$^{-1}$. 
It is also worth mentioning 
that starting from the year 2002 all the P Cygni profiles measured in the spectrum of V838 Mon have 
had emission peaks at heliocentric velocities $80-100$\,km\,s$^{-1}$ \citep{2007ASPC..363...13M, 
2009ApJS..182...33K, 2021A&A...655A..32K}, and under the assumption of spherical winds, the 
position of the emission peak indicates the radial velocity of the star. 
We note here that in the light of the recent ALMA observations \citep{2021A&A...655A..32K}, 
which resolve the position 
of both components in V838 Mon, it is clear that the maximum flux emission 
is located at the position of the RSG (see their Fig. 3). 
Therefore, the  above-mentioned radial velocity 
 represents the velocity of the RSG, rather than the true barycentric (i.e. systemic) velocity of the system.

\section{Conclusions}\label{S-conc}

We present new extensive and homogeneous observational datasets of 
the intriguing object V838 Mon.
Before calibrating the photometric data we   performed a solid
stability check of the comparison stars.
Our photometric and spectroscopic monitoring of this peculiar target over more than a decade 
reveals important insights into its temporal evolution. We note a brightening in photometry, in 
particular in the blue bands, and a simultaneous weakening of the molecular absorption bands. 
The resonance lines of Ca\,{\sc i} and Li\,{\sc i} consist of two absorption components, a 
blue-shifted one tracing the expanding shell and a red-shifted one originating from the ongoing but 
steadily weakening wind from the SG. 
The continuous strengthening of the H$\alpha$ emission traces the 
gentle reappearance of the hot companion.
Moreover, the detection of the persistent Ti\,{\sc i} line
at about $+90$\,km\,s$^{-1}$ refers to the radial velocity of the RSG component 
rather than matter infall as previously proposed.

We note that all the observed changes discussed in Sect.~\ref{S-disc}  
may be the result of multiple, combined effects. An illustrative example is the
observed gradual increase in continuum flux in the blue, which most likely is caused 
by the interplay of (i) an increase in effective temperature of the SG star, 
(ii) a drop in the extinction by the circumstellar dust around both the SG 
and the hot companion, and (iii) a weakening of the SG star's wind. How much each of these effects contributes requires comprehensive modelling of 
combined, high-quality photometric and high-resolution spectroscopic data. 
While such a thorough analysis is beyond the  reach of our spectroscopic data, 
our results can serve at least as a homogeneous set of observational constraints 
for future modelling of the challenging binary system V838 Mon.

\begin{acknowledgements}

We thank the anonymous referees for the constructive suggestions that helped to 
improve the manuscript.
Based on observations made with the Nordic Optical Telescope, owned in collaboration by 
the University of Turku and Aarhus University, and operated jointly by Aarhus University, 
the University of Turku and the University of Oslo, representing Denmark, Finland and Norway, 
the University of Iceland and Stockholm University at the Observatorio del Roque de los 
Muchachos, La Palma, Spain, of the Instituto de Astrofisica de Canarias.
The data presented here were obtained with ALFOSC, which is provided by the 
Instituto de Astrofisica de Andalucia (IAA) under a joint agreement with the 
University of Copenhagen and NOT. Observations on 2019 November 29 were done as 
part of the NOT science school for upper secondary school students organised in Tuorla Observatory. 
In addition, this paper uses 
observations made at the South African Astronomical Observatory (SAAO).

Analyses done in this paper use a GAIA shoftware. GAIA is a derivative of the Skycat 
catalogue and image display tool, developed as part of the VLT project at ESO. 
Skycat and GAIA are free software under the terms of the GNU copyright. 
The 3D facilities in GAIA use the VTK library. For further details, 
see http://starlink.eao.hawaii.edu/starlink

T.L., M.K., and B.D. acknowledge financial support from the Czech Science Foundation
(GA \v{C}R, grant numbers 20-00150S and 19-18647S). The Astronomical Institute of the
Czech Academy of Sciences is supported by the project RVO:67985815.
This project has received funding from the European Union's
Framework Programme for Research and Innovation Horizon 2020 (2014-2020)
under the Marie Sk\l{}odowska-Curie Grant Agreement No. 823734.

All the data from the NOT can be acquired through the NOT FITS Header  
Archive\footnote{http://www.not.iac.es/observing/forms/fitsarchive/}. 
The rest of the data can be obtained from TL on a reasonable request.  
\end{acknowledgements}


\bibliographystyle{aa}
\bibliography{literature838}


\begin{appendix}

\section{Complementary optical photometry}\label{A-phot}

During the years 2002--2008 photometric observations of V838 Mon were carried out at Tartu Observatory (TO)
 with the \hbox{0.6 m} ZEISS telescope and a thermoelectrically cooled CCD camera HPC. 
Filters $BVR_CI_C$ in the Johnson-Cousins system were used. 
Starting from 2013, an Andor Ikon-L (IKON)
with an Optec Inc. Bessell filter set has been used. The  FOV of both cameras is $13'\times13'$. 
During these years, data from 48 observing nights were collected. 

In addition, starting from 2014, the 31.4 cm telescope Planewave CDK 12.5 (also known as RAITS) in TO 
was used to obtain photometric data of V838 Mon on 24 nights. 
The telescope is equipped with a CCD camera consisting of a 
Apogee Alta U42 camera 
and a set of second-generation Astrodon Bessell $BVR_CI_C$ filters. The FOV  
of that telescope is $38'\times38'$.

From 2007 January 31 to March 13 higher cadence observations were secured at the
South African Astronomical Observatory (SAAO). A 1.0 m telescope was exploited, 
equipped with a STE4 CCD camera 
(FOV $5\farcm 3\times 5\farcm 3$)
and $UBVR_CI_C$ filters. 
In total 22 photometric measurements were acquired. 

\section{Aligning datasets from different telescopes}\label{A-align}

\begin{figure}[!h]
\centering
\includegraphics[width=9cm]{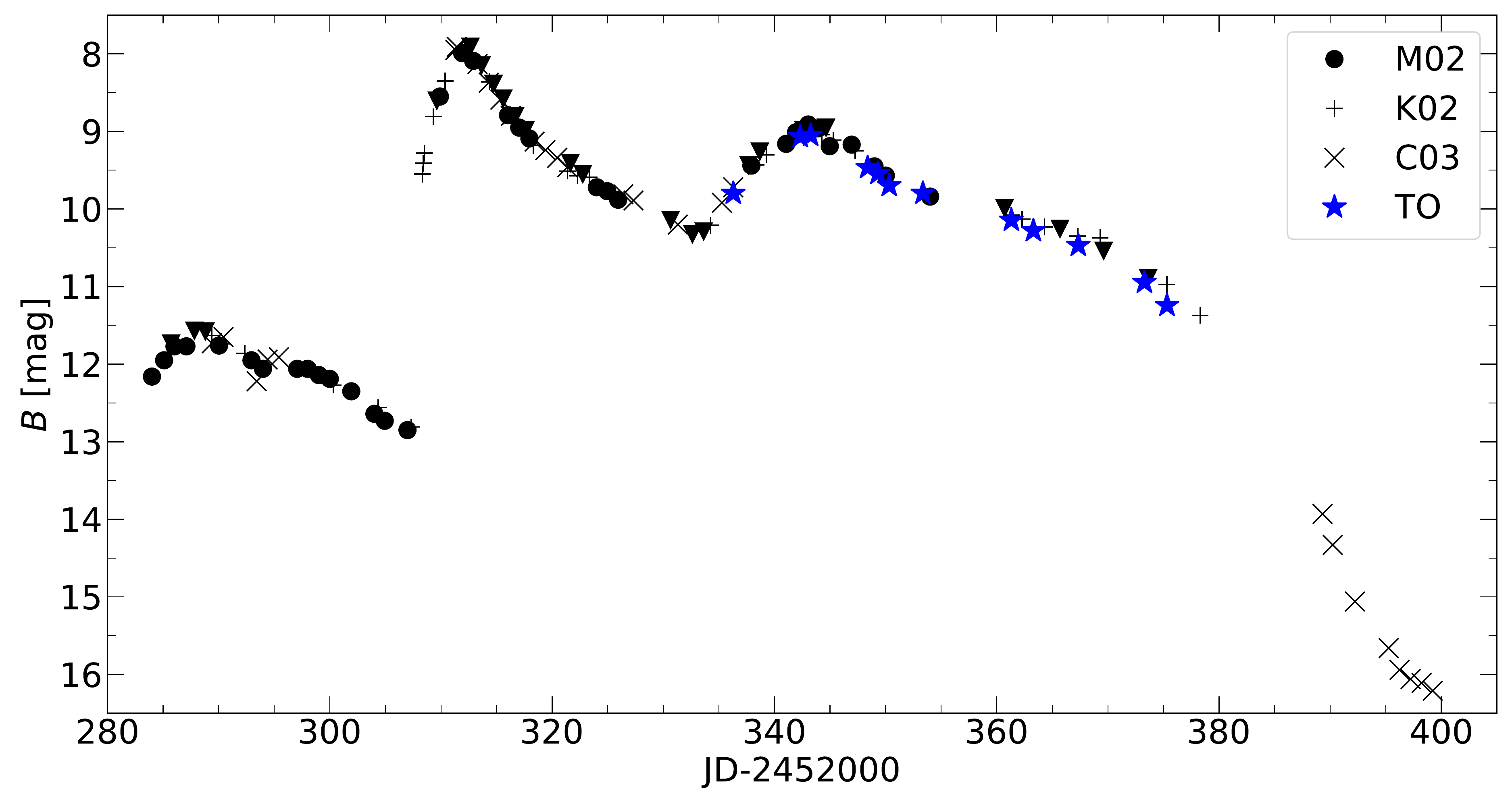}
\caption{
V838 Mon outburst between January and May 2002 in the $B$ band, as reported by various authors.
A good match between the literature (black) and our (blue) values is visible.
M02 refers to data from \cite{2002A&A...389L..51M}, 
K02 to \cite{2002MNRAS.336L..43K}, C03 data to \cite{2003MNRAS.341..785C}, and TO to data from this work.}
\label{F-lc2002B}
\end{figure}

\begin{figure}[!h]
\centering
\includegraphics[width=9cm]{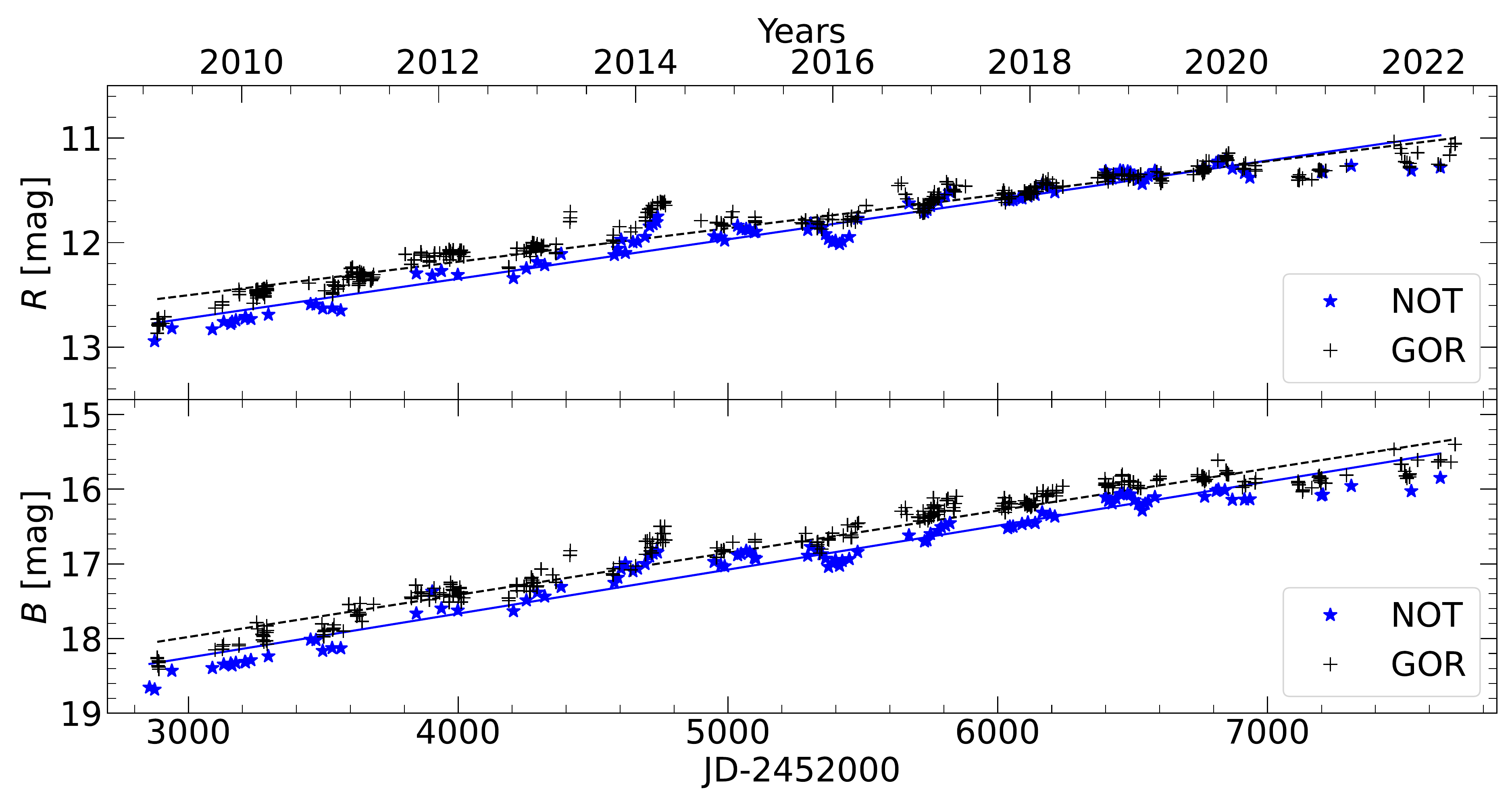}
\caption{
Demonstration of the calibration issues in the $R$-band (top) and $B$-band (bottom) 
  magnitudes between our NOT (blue) and the GOR20 (black) values.
See text for more details. }
\label{F-lc2009B}
\end{figure}

As pointed out by \cite{2007ASPC..363....3H}, in early 2002 it was easy to align 
light curves obtained with different instruments, while later, due to the extreme colours of V838 Mon, 
no comparison stars with matching colours have been available, and hence matching has not been trivial. 
The offsets between published photometric datasets are also discussed in GOR20. 
As an example, we plot in Fig.~\ref{F-lc2002B} our TO data from 2002 
(in blue) together with literature values 
(in black). A reasonably good match can be seen between the  datasets. 
In Fig.~\ref{F-lc2009B} we plot our $BR$ magnitudes 
taken with  NOT starting from 2009 together with the $BR$ data of GOR20 
collected during the same period.\footnote{http://www.vgoranskij.net/v838mon.ne3}
A straight line is fitted through each dataset. 
As can be seen, the fitted lines in the $B$ filter can be considered parallel, 
implying a constant offset of about 0.4 mag 
between our NOT and GOR20 standard magnitudes. In the $R$ band the fitted lines are 
not parallel, probably due to the above-mentioned calibration difficulties.  
The issue of offsets changing over time was also pointed out by GOR20.

Similar offsets were found between our NOT  
and some of the TO data starting from 2008. This is especially obvious 
for the $B$ data from RAITS (black points in Fig.~\ref{F-lcour}) taken in 2014 and 2015, 
which in addition to the calibration difficulties might also result 
from the rather small telescope size and from the faintness of 
the V838 Mon during the past ten years in bluer bands. 
On the other hand, the possible reason for the small systematic offset 
may be related to the exact position and shape of filter transmittance curve 
relative to the sharp-edged spectral features of V838 Mon.

However, because our measurements fit well with other published data during the 2002 outburst, 
we consider our calculated standard magnitudes to be  reliable.
Moreover, we   restrict our analysis of the 
light curve from 2009 onwards to the homogeneous data collected with  NOT 
because these data were obtained with a uniform telescope and instrument setting.

\end{appendix}

\end{document}